\documentclass[prl,aps,showpacs,twocolumn,preprintnumbers,amsmath,amssymb,superscriptaddress,longbibliography,nofootinbib]{revtex4-2}
\usepackage[english]{babel}
\usepackage{amsmath,amssymb,bbm,mathrsfs,mathtools,multirow,diagbox,xcolor,graphicx,tabularx,comment,amsfonts,dsfont,units,comment}
\usepackage[english]{babel}
\usepackage{amsmath,amssymb,bbm,mathrsfs,bm,braket,color,graphicx,comment,amsfonts,dsfont}
\usepackage[colorlinks,linkcolor=blue,citecolor=blue,urlcolor=blue]{hyperref}
\usepackage[mathscr]{euscript}
\usepackage{physics}
\usepackage{xcolor}
\usepackage{bm}
\usepackage{orcidlink}
\usepackage[normalem]{ulem} 

\makeatletter
\def\maketitle{
\@author@finish
\title@column\titleblock@produce
\suppressfloats[t]}
\makeatother

\begin{document}
\title{Topological diffusive metal in amorphous transition metal monosilicides}

\author{Selma Franca\orcidlink{0000-0002-0584-2202}}
\email{selma.franca@neel.cnrs.fr}
\author{Adolfo G. Grushin\orcidlink{0000-0001-7678-7100}}
\email{adolfo.grushin@neel.cnrs.fr}
\affiliation{Univ. Grenoble Alpes, CNRS, Grenoble INP, Institut Néel, 38000 Grenoble, France}

\date{\today}
\begin{abstract}
In chiral crystals crystalline symmetries can protect multifold fermions, pseudo-relativistic masless quasiparticles that have no high-energy counterparts. 
Their realization in transition metal monosilicides has exemplified their intriguing physical properties, such as long Fermi arc surface states and unusual optical responses. 
Recent experimental studies on amorphous transition metal monosilicides suggest that topological properties may survive beyond crystals, even though theoretical evidence is lacking.
Motivated by these findings, we theoretically study a tight-binding model of amorphous transition metal monosilicides.
We find that topological properties of multifold fermions survive in the presence of structural disorder that converts the semimetal into a diffusive metal.
We characterize this topological diffusive metal phase with the spectral localizer, a real-space topological indicator that we show can signal multifold fermions.
Our findings showcase how topological properties can survive in disordered metals, and how they can be uncovered using the spectral localizer.

\end{abstract}
\maketitle

\textit{\textcolor{blue}{Introduction}} ---
Crystalline topological metals host quasiparticles classified according to 
the symmetries required to protect them.
For example, Weyl semimetals require no symmetries to realize Weyl quasiparticles,
which are spin-half, gapless low-energy quasiparticles governed by the Weyl equation~\cite{Armitage2018}.
Weyl bands disperse linearly around a two-band crossing point, 
accompanied by a quantized flux of Berry curvature, known as the monopole charge. 
The absence of symmetry requirements endows Weyl points with a relative 
robustness against disorder~\cite{Fradkin1986, Fradkin1986a, Altland2015, Syzranov2016, Pixley2016a, Altland2016, Sbierski2016, Louvet2016, Louvet2017, Luo2018, Balog2018, Pixley2021, Nandkishore2014, Pixley2015, Pixley2016, Gurarie2017, Buchhold2018, Buchhold2018a, Wilson2020}, 
explaining why they have been predicted to survive even in non-crystalline lattices~\cite{Yang2019}.

Higher-spin generalizations of Weyl quasiparticles known as multifold fermions, 
predicted and observed in chiral crystals~\cite{Manes2012,Bradlyn2016, Tang2017, Chang2018, Sanchez2019, Takane2019, Rao2019, Schroter2019, Wu2019, Ni2021},  seem more delicate.
Their bands disperse linearly around a multi-band crossing points and can have an associated monopole charge.
However, in contrast to Weyl quasiparticles, they require crystalline symmetries to ensure their robustness.
The effect of disorder on multifold semimetals is much less explored~\cite{Hsu2022, Kikuchi2023}, and it seems
paradoxical that topology can survive the absence of long-range lattice order.

In this work we investigate to what extent the above expectation holds in a non-crystalline amorphous model.
Our main result is that topological properties of multifold fermions can survive the absence of crystal symmetry.
Recently, amorphous insulators have been predicted and observed to display topological phases, 
owing to the finite energy scale endowed by the gap~\cite{Agarwala2017, Mansha2017, Mitchell2018, Chern2019, Marsal2020, Sahlberg2020, Ivaki2020, Costa2019,  Focassio2021, Wang2022, Ma2022, Mano2019, Mukati2020, Corbae2023a}.
Indeed, models of amorphous Chern insulators~\cite{Agarwala2017, Mansha2017, Mitchell2018, Marsal2020, Sahlberg2020, Ivaki2020}, 
quantum spin-Hall insulators~\cite{Costa2019, Focassio2021, Wang2022, Ma2022} and 3D topological insulators~\cite{Agarwala2017, Mano2019, Mukati2020}, demonstrate that topology 
survives the amorphicity, and can even be induced by it~\cite{Wang2022}.
Moreover, 
average crystalline symmetries can also protect amorphous topological states, 
provided that the disorder strength is smaller than the band gap~\cite{Spring2021, Agarwala2020, Wang2021, Peng2022, Tao2023}.

In turn, the survival of topology in 
amorphous metals is much more challenging to address due to the absence of a gap.
Methods to detect metallic topology in real-space are scarce, 
especially in the presence of time-reversal symmetry 
where local Chern markers~\cite{Loring_2010,Bianco2011} are identically zero.

To make progress, here we amorphisize a known crystalline model of 
a chiral crystal in space group 198~\cite{Chang2017, Tang2017, Pshenay-Severin2018, Takane2019}. 
Materials in this space group, such as the 
transition metal monosilicides RhSi or CoSi, lack inversion and mirror symmetries yet exhibit nonsymmorphic
symmetries.
These materials manifest exotic physical properties such as multifold fermions 
at the Fermi level, long Fermi arcs surface states~\cite{Chang2017}, 
a quantized circular photogalvanic effect~\cite{deJuan17,Chang2017,Flicker2018, Rees2020, Ni2020, Ni2021} 
and unusual magneto-transport features~\cite{Bradlyn2016, Chang2017, Guo2022, Hu2023}. 
Moreover, a recent experimental study on amorphous CoSi (a-CoSi) 
has found a range of intriguing magneto-transport properties~\cite{Molinari2023}.

Using the recently introduced spectral localizer
~\cite{Loring2015, Loring2019, Loring2018, Schulz-Baldes2021, Schulz-Baldes2022, Cerjan2022}, 
we find that multifold fermions enter a topological diffusive metal (TDM) phase in 
the presence of moderate structural disorder.
We find that localizer in-gap modes can be traced back to the existence of multifold fermions
and coexisting with spectral properties characteristic of a diffusive metal~\cite{Pixley2021}.
Upon increasing disorder, the localizer in-gap modes are lost, leaving behind a trivial diffusive metal (DM)
that eventually localizes into a trivial Anderson insulator (AI).
Using the spectral localizer to define TDMs can be 
extended to any symmetry class, and hence is the main result of this work
(see Fig.~\ref{fig:fig1}a).

\textit{\textcolor{blue}{Model Hamiltonian}} --- 
Amorphous systems lack long-range order, but they display short-range ordering, dictated by the local chemistry of the elements~\cite{Zallen}.
This implies the existence of preferred bond lengths and angles peaked around the crystalline values~\cite{Toh2020,Corbae2023,Ciocys2023}.
Hence, we first 
revisit the crystalline model of RhSi and CoSi, in space group 
$198$, on which we base our amorphous model.
This space group has three non-intersecting twofold screw symmetries 
$s_{2x,y,z}$ and a diagonal cubic threefold rotation $C_{3,111}$. 
The spin-orbit coupling in RhSi~\cite{Chang2017} and CoSi~\cite{Xu2020} 
is weak (tens of meV), and is neglected in our simulations.
In this case, the band structure near the Fermi level is captured by 
a tight-binding Hamiltonian with four $s$-type orbitals (A, B, C, D) 
positioned at $(0,0,0), (\frac{a}{2},0,\frac{a}{2}), (\frac{a}{2}, \frac{a}{2},0)$ 
and $(0,\frac{a}{2}, \frac{a}{2})$, 
see Fig.~\ref{fig:fig1}b~\cite{Chang2017}.
In the following, we measure all distances in units of $a$. 
Nearest-neighboring orbitals are connected 
by inter-orbital hoppings while second nearest-neighbors 
are connected via intra-orbital hoppings. 
Fig.~\ref{fig:fig1}c, illustrates the inter-orbital hoppings within the unit cell, 
which take two values $(v_1\pm v_p)/4$, depending on the bond orientation.
The amplitude $v_2/2$ of intra-orbital hoppings is independent of the bond orientation. 
The Bloch Hamiltonian is discussed further in the Supp. Mat. (SM)~\cite{SM}.

It is convenient to consider two parameter regimes, expressed in eV: (1) when only $v_p=-0.762$ is non-zero,
and (2) when $v_1 = 0.55, v_2 = 0.16, v_p = -0.762$ eV.
The hopping amplitudes in regime (2) 
are chosen such that the crystalline tight-binding Hamiltonian~\cite{SM} reproduces well 
the density-functional theory calculated band structure 
of RhSi near the Fermi level~\cite{Chang2017}.
Hence, in the following we refer to regime (2) as a-RhSi regime.
The red curves in Figs.~\ref{fig:fig1}d,e 
show bulk spectra for the two regimes, respectively. 
In regime (1), the spectrum is doubly degenerate in the entire Brillouin zone 
and features two double-Weyl fermions,
one at $\Gamma$ and one at the $R$ point, occurring at the same energy $E=0$. 
In the a-RhSi regime, $v_1, v_2$ turn the double-Weyl at $\Gamma$
into a threefold fermion, energetically shifted with respect 
to the double-Weyl fermion at the $R$ point.
In both regimes, the crossings at $\Gamma$ and $R$ 
have monopole charges $C=2$, and $-2$, respectively.

\begin{figure}[tb]
\includegraphics[width=8.6cm]{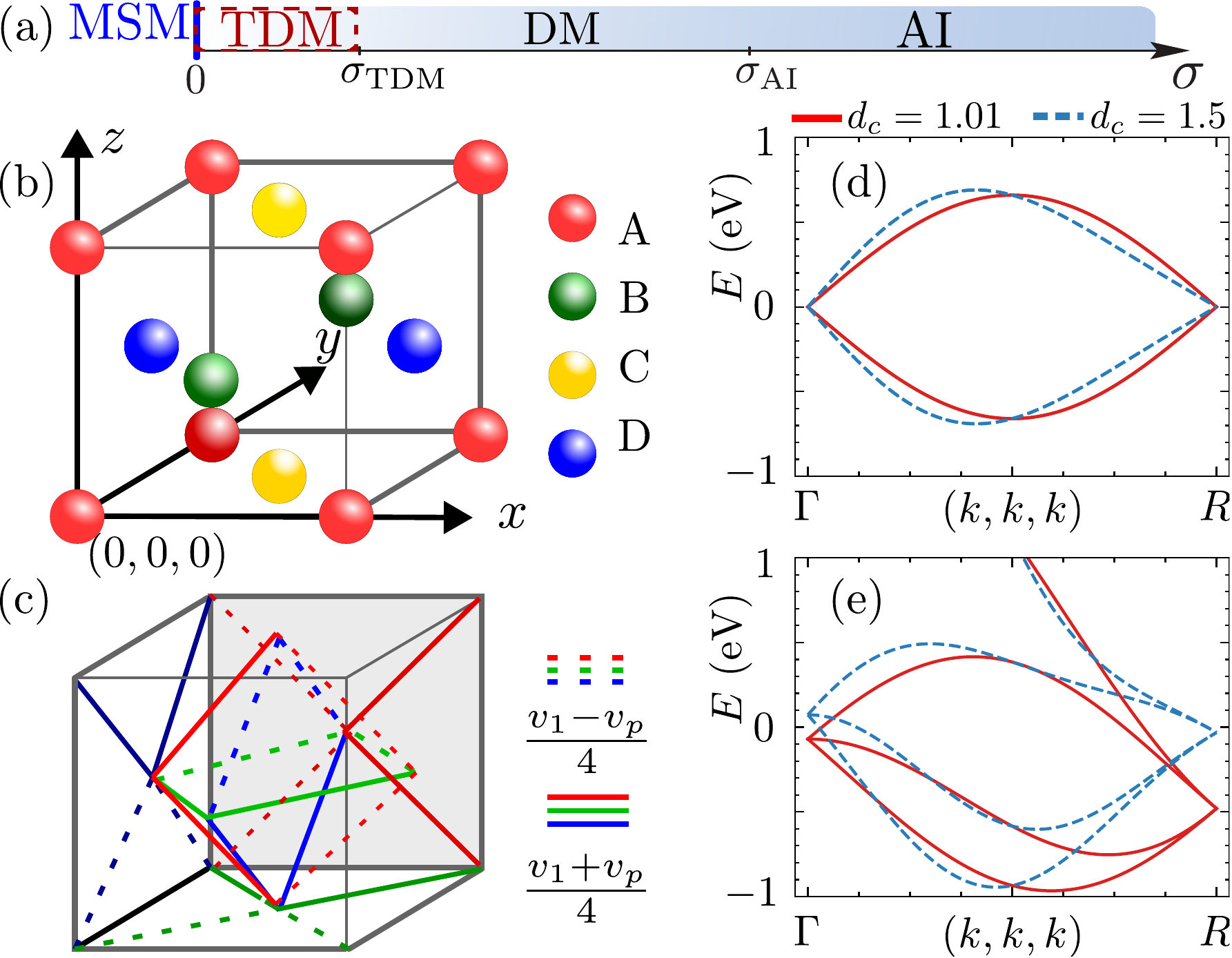}
\caption{
(a) Schematic phase diagram found in this work, including the multifold semimetal (MSM),
topological diffusive metal (TDM), the diffusive metal (DM) and the Anderson insulator (AI) phases, 
as a function of the disorder variance $\sigma$.
The TDM persists until $\sigma_{\mathrm{TDM}}$ that is defined in Fig.~\ref{fig:fig4}, and is signaled by in-gap states of the localizer as well as a finite DOS at $E_F$ typical of a diffusive metal. 
The trivial DM phase is delimited by $\sigma_\mathrm{AI}$, defined in Fig.~\ref{fig:fig3}, from the AI phase.
(b) Orbitals of the crystalline unit cell.  
(c) The nearest-neighbor inter-orbital hoppings 
$\frac{v_1\pm v_p}{4}$ represented by solid and dashed lines. 
The color denotes hoppings between different set of orbitals: 
red for A-B/C-D, blue for A-D/B-C and green for A-C/B-D.
(d) Band structure for parameter regime (1) with $v_1= v_2=0, v_p = -0.762$,
(e) Band structure for parameter regime (2), a-RhSi, with $v_1=0.55,v_2=0.16,v_p = -0.762$.
Solid and dashed curves correspond to maximum hopping radii $d_c=1.01,1.5$, respectively.
In (d), the spectrum is doubly degenerate with two double-Weyl fermions occurring at $\Gamma$ and $R$ points.
In (e), a-RhSi regime, we see a threefold fermion at $\Gamma$ point and a double-Weyl fermion at $R$ point.
}
\label{fig:fig1}
\end{figure}

We create the amorphous lattice by displacing every site $n$ 
(representing a single orbital) of crystalline RhSi by
$\delta \mathbf{r}_n = (\delta x_n, \delta y_n, \delta z_n)$ 
drawn from a Gaussian distribution 
\begin{equation}\label{eq:gdist}
    D(\delta \mathbf{r}_n) = \frac{1}{2\pi \sigma^2} \exp[-\frac{|\delta \mathbf{r}_n|^2}{2\sigma^2}].
    \end{equation}
The variance $\sigma^2$ is typically proportional to the quenching temperature 
to form the amorphous solid, $\sigma^2 \propto k_B T$~\cite{Wang2022}.
To avoid artificial clustering of sites,
we impose a minimal distance of $d_\mathrm{min}=0.4$~\cite{Spring2021}.
The possible hoppings $\tilde{v}_{\alpha}$ ($\alpha = 1,2,p$) 
between sites at positions 
$\mathbf{r}_{n} = (x_n,y_n, z_n)$ and $\mathbf{r}_{m}=(x_m, y_m, z_m)$ 
are determined from
\begin{equation}~\label{eq:hopdef1}
\tilde{v}_{\alpha} = \tilde{v}_{\alpha} (d) \tilde{v}_{\alpha} (\theta, \phi) \exp[1-\frac{d}{d_{\alpha} ^0}] \Theta(d_c-d),
\end{equation}
in spherical coordinates $(d,\theta,\phi)$ with $d= |\mathbf{r}_n-\mathbf{r}_m|$.
Here, $d_{\alpha} ^0$ depends on whether the hopping is 
inter-orbital, where $d_{1} ^0 = d_{p} ^0=1/\sqrt{2}$,
or intra-orbital, where $d_{2} ^0 = 1$.

Since the intra-orbital hopping $v_2/2$ of crystalline RhSi 
is independent of bond orientation~\cite{SM},
we take $\tilde{v}_2(\theta, \phi)=1$ and $\tilde{v}_2 (d) = v_2/d$. 
The inter-orbital hopping of crystalline RhSi has amplitudes $(v_1 \pm v_p)/4$, 
see Fig.~\ref{fig:fig1}c, 
where the hopping $v_p$ is direction dependent, unlike $v_1$. 
Hence, we take $\tilde{v}_1(\theta, \phi)=1$, $\tilde{v}_1 (d) = v_1/\sqrt{2} d$ 
and $\tilde{v}_p (d) = v_p/d$.
In contrast, the hopping $\tilde{v}_p (\theta, \phi)$ depends on 
the type of orbitals that form the bond as detailed in the SM~\cite{SM}. 
The $\tilde{v}_{\alpha}$ recover the original 
tight-binding Hamiltonian in the crystalline limit when
$\sigma \rightarrow 0$~\cite{Chang2017, SM}.

The step function $\Theta$ in Eq.~\eqref{eq:hopdef1} 
ensures that the maximum distance between two sites is $d_c$,
whose effect is shown in Figs.~\ref{fig:fig1}d and e. 
Notably, for a-RhSi longer range hoppings reduce the energy difference between threefold and double Weyl fermions.
In our simulations, $d_c = 1.5$ which allows to account for longer-range hoppings (see SM~\cite{SM}).

Lastly, to account for possible potential disorder~\cite{Spring2021}, we add  
random onsite potentials drawn from the Gaussian distribution Eq.~\eqref{eq:gdist}.
Thus, our model accounts for all types of disorder expected in amorphous solids: 
on-site, hopping and structural disorder.

\textit{\textcolor{blue}{Spectral properties} }--
The density of states (DOS) characterizes disordered topological semimetals~\cite{Pixley2015}, 
and can be defined as $\rho^{\lambda} (E) = \frac{1}{V} \sum_m \delta(E-E_m)$,
where $\lambda$ labels a disorder realization, 
$m$ runs over all states of the system,
and $V = 4 L^3$ for a cubic system with $L$ unit cells in each direction.
For every disorder realization, the DOS is calculated using 
the numerically efficient kernel polynomial method (KPM),
which relies on a Chebyshev polynomial expansion up to order $N_C$~\cite{Weisse2006}.
In the following, we study the disorder averaged DOS
$\bar{\rho} (E) = \frac{1}{\rm N_{dis}} \sum_{\lambda=1}^{\rm N_{dis}} \rho^{\lambda} (E)$ 
with $\rm N_{dis} = 16$.

We focus first on regime (1) 
that has two double Weyl fermions at $E= 0$ in the crystalline limit. 
Figs.~\ref{fig:fig2}a and b show $\bar{\rho}(E)$ 
for different disorder strengths and $\bar{\rho}_0 \equiv \bar{\rho} (E=0) $ 
for different KPM orders $N_C$, respectively.
We see that for $\sigma \lessapprox 0.04$, $\bar{\rho}(E) \to |E|^2$,
as in periodic systems with (double-) Weyl 
fermions at the same energy.  
Once the disorder strength is increased up to
$\sigma\approx 0.07$, $\bar{\rho} \propto |E|$ close to $E=0$.
Additionally, at $\sigma = 0.1$ the DOS at $E=0$ becomes nonzero. 
Fig.~\ref{fig:fig2}b reveals that $\bar{\rho}_0 \neq 0$ for $\sigma \geq 0.1$ 
signaling that the system becomes a diffusive metal, 
a phase with constant DOS in a range of energies.

 \begin{figure}[t]
\includegraphics[width=8.6cm]{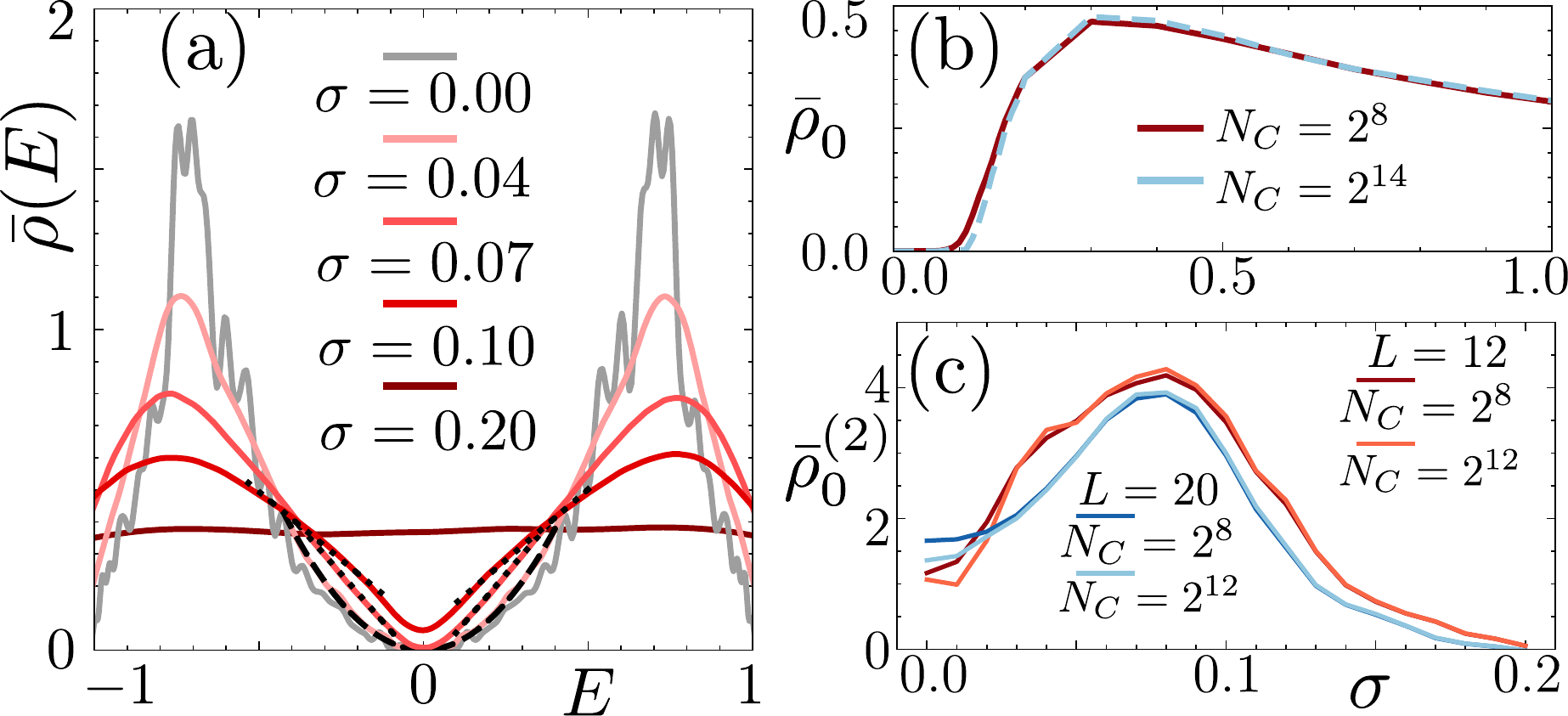}
\caption{
(a) $\rho(E)$ vs $E$ as a function of $\sigma$. 
Dashed and dotted lines represent fit functions $\alpha E^2$ and $\alpha + \beta |E|$, respectively. 
(b) The averaged zero-energy DOS $\bar{\rho}_0$ as a function of disorder strength $\sigma$; 
$\rho_0$ becomes nonzero at $\sigma \approx 0.1$.
(c) $\bar{\rho}^{(2)} (0)$ vs $\sigma$ peaks around $\sigma_c = 0.08$ independent of system size ($L=12,20)$ and KPM order $N_C$.
We define $\bar{\rho}^{(2)} (0)=\sum_{\lambda=1}^{N_{\rm dis}}(\rho^{\lambda})^{(2)} (0) $ where $(\rho^{\lambda})^{(2)} (0) $ is estimated from a fit $\rho^{\lambda}(E) = \rho^{\lambda}_0+ (\rho^{\lambda})^{(2)} (0) E^2$ 
in the energy range $(-0.2,0.2)$ for independent disorder realizations $\lambda$.}
\label{fig:fig2}
\end{figure}

This behavior suggests a putative quantum critical point (QCP) at a certain $\sigma_{c}$,
where the semimetal phase is replaced by a diffusive metal~\cite{Pixley2016, Pixley2021}.
To study this phase transition in more detail, Fig.~\ref{fig:fig2}c shows 
$\bar{\rho}_0^{(2)}$, extracted from a low energy fit
$\bar{\rho}(E) = \bar{\rho}_0 +  \bar{\rho}_0^{(2)} E^2$ to the DOS~\cite{Pixley2021}.
It remains finite with a maximum at $\sigma_c \approx 0.08$, that shifts little with increasing system 
size or the order of the KPM expansion.
This non-divergent behavior signals that the putative QCP is avoided~\cite{Pixley2021}.

Avoiding such QCP is enabled by the presence of 
statistically-rare states~\cite{Pixley2021}.
Rare states are low-energy eigenstates that are quasi-bounded to 
the real space regions with uncharacteristically 
large potential strengths that are statistically rare.
In the thermodynamic limit, these statistically rare events are likely to 
occur for any nonzero $\sigma$.
As a result, $\bar{\rho}_0$ becomes exponentially small in disorder strength but nonzero, 
implying a crossover from semimetal to diffusive 
metal phase instead of a perturbative transition~\cite{Pixley2016, Pixley2021},     
see SM~\cite{SM} for more details.

Importantly, it is the vanishing DOS at the band 
crossing that makes rare-states dominate the physics of disordered Weyl semimetals. 
The DOS vanishes at $E=0$ for parameter regime (1), where two Weyl nodes 
coexist at the Fermi level, but not for the a-RhSi regime (2).
In the latter, disorder can couple states without energy penalty~\cite{Trescher2017}, 
turning the a-RhSi regime into a diffusive metal for any infinitesimal amount of disorder.

\textit{\textcolor{blue}{Anderson localization} }--
While eigenstates of diffusive metals are extended, 
sufficiently strong disorder will turn metallic systems into 
Anderson insulators with localized eigenstates~\cite{Pixley2015}.
Localized states interact weakly, thus producing an uncorrelated energy spectrum
that obeys a Poisson distribution function~\cite{Akker-Mont}.
On the metallic side, the overlap of delocalized states leads to 
the repulsion of associated energy levels. 
For spinless and time-reversal symmetric systems, 
like a-RhSi, such a spectrum falls under the Gaussian Orthogonal Ensemble (GOE) 
of random matrices~\cite{Beenakker1997}.
To distinguish between metallic and insulating regimes, 
we calculate the adjacent energy level spacing ratio and the 
inverse participation ratio (IPR) of states at the Fermi level $E_F$.
The adjacent level spacing ratio 
is defined as
\begin{equation}
r = \frac{1}{N_E-2} \sum_m \frac{\min(\tilde{E}_{m,m-1}, \tilde{E}_{m+1, m})}{\max(\tilde{E}_{m,m-1}, \tilde{E}_{m+1,m})},
\end{equation}
where $\tilde{E}_{m,n} = E_m-E_{n}$ and the energy levels are arranged such that $E_m > E_{m-1}$.
The sum is performed over $N_E$ energy levels within the interval $[ E_F-\Delta E, E_F+\Delta E ]$.
The GOE and Poisson spectra have $r_{\rm GOE} \approx 0.54$ and 
$r_{\rm P} \approx 0.39$, respectively~\cite{Atas2013}.
To quantify the localization of a set of eigenstates near $E_F$, we use the
IPR defined as ${\rm{IPR}} = \sum_m \sum_{\mathbf{r}_n} |\Psi_m (\mathbf{r}_n)|^4/N_E$.
Here $\Psi_m$ is the eigenstate corresponding to $m$-th eigenvalue, 
and the sum is taken over the same energy window $[ E_F-\Delta E, E_F+\Delta E ]$ as for $r$. 
The IPR is close to zero (unity) for delocalized (localized) states. 

\begin{figure}[t]
\includegraphics[width=1\columnwidth]{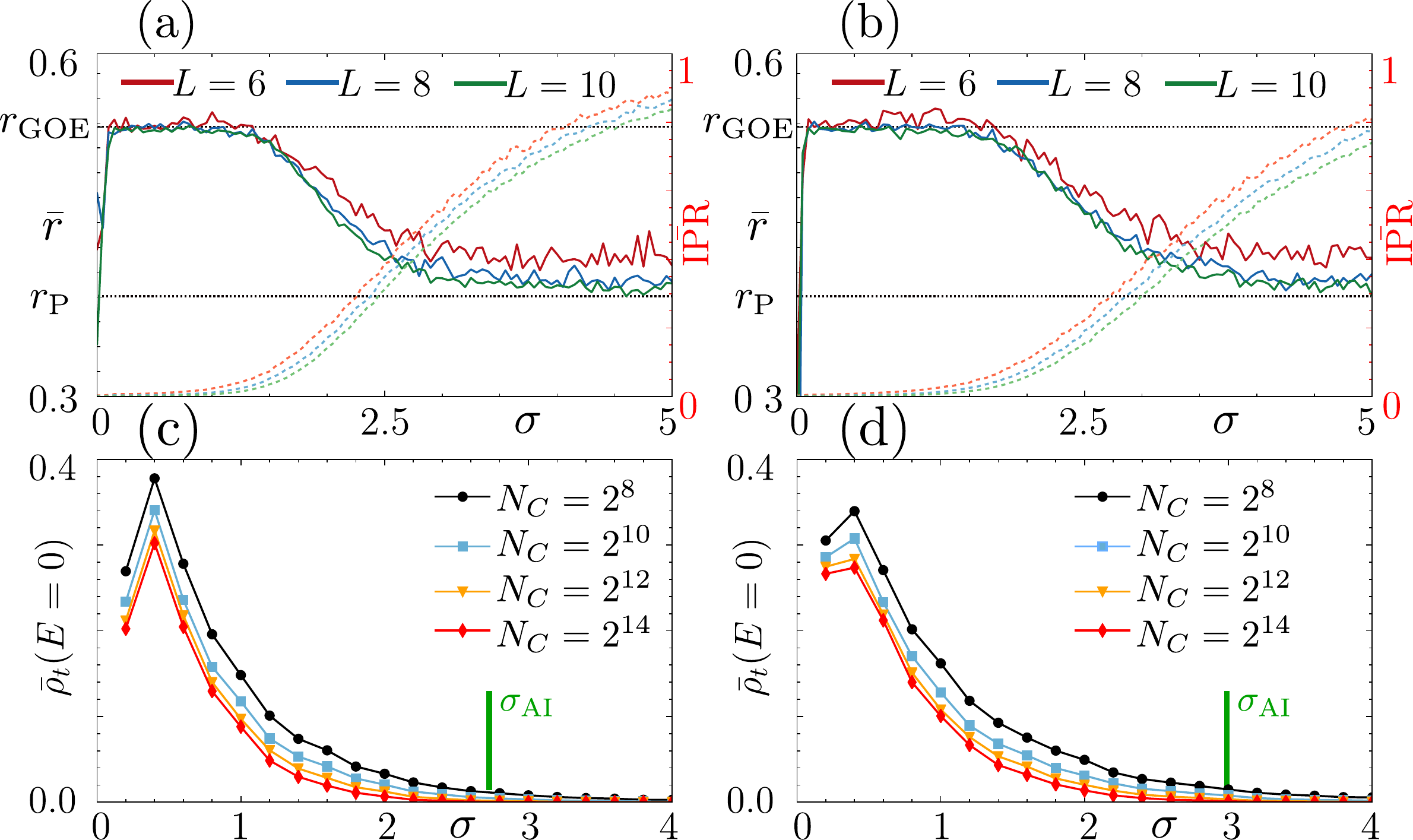}
\caption{ 
(a) and (b) show disorder averaged adjacent level spacing ratios (solid lines) and inverse participation ratios (dashed lines) at $E_F = 0$ as a function of disorder strength $\sigma$ for a system in parameter regime (1) and a-RhSi regime, respectively. 
(c) and (d) show the disorder averaged typical DOS for different orders of KPM expansion $N_C$ as a function of $\sigma$ for a system in parameter regime (1) and a-RhSi regime, respectively.
Here, $\sigma_{\rm AI}$ is the critical disorder strength at which the system transitions from a DM to an AI phase. 
}
\label{fig:fig3}
\end{figure}

In the following, we focus on small system sizes $L=6,8,10$ where exact diagonalization is possible.
We fix $E_F = 0, \Delta E = 0.1$ in order to probe the physics near the multifold crossings, and calculate the disorder averaged 
$r$ and IPR: $\bar{r} \; (\rm{\bar{IPR}}) = \frac{1}{\rm N_{dis}} \sum_{\lambda=1}^{\rm N_{dis}} r^{\lambda} \; (\rm IPR^{\lambda})$, where $\rm N_{dis} = 101$.
The results are shown in Fig.~\ref{fig:fig3}a,b for the parameter regime (1) and the a-RhSi regime, respectively.
In both cases, we observe that the transition from $\bar{r} \approx 0.54$ (GOE) to $\bar{r} \approx 0.39$ (Poisson) occurs gradually due to finite size effects, a behavior also reflected in the $\rm{\bar{IPR}}$ that changes from $0$ to $\approx 1$ as $\sigma$ is increased.

To find the disorder strength for which the topological phase transition into Anderson insulator phase occurs, we use the typical DOS at $E = 0$ defined as 
$\bar{\rho}_0^t= \exp[\sum_{\mathbf{r}_n}  (\sum_{\lambda=1}^{N_{\rm dis}} \log \rho_0^{\lambda} (\mathbf{r}_n)/N_{\rm dis}) / V_{t} ]$.
Here, $\rho_0^{\lambda} (\mathbf{r}_n)$ is the local DOS at site $\mathbf{r}_n$ and energy $E=0$, while $V_t = 32$ is the total number of sites ($8$ sites in the bulk for each orbital type).
We calculate the typical DOS for a system of linear size $L=20$.
Since the typical DOS is not a self-averaging quantity, we set the number of disorder realizations to $N_{\rm dis} = 51$.
The results are shown in Fig.~\ref{fig:fig3}c and  Fig.~\ref{fig:fig3}d for regime (1) and the a-RhSi regime, respectively.
The typical DOS is sensitive in the order of the KPM expansion $N_C$~\cite{Pixley2015}, and to determine $\sigma_{\rm AI}$ we extrapolate $\rho_0^t$ to zero using the Richardson extrapolation method for data points obtained with $N_C = 2^{14}$.
We obtain $\sigma_{\rm AI} = 2.715$ for regime (1) and $\sigma_{\rm AI} = 2.985$ for the a-RhSi regime.
%

\textit{\textcolor{blue}{Topological phase diagram} }--
Lastly, we are interested in quantifying to what extent 
the topological properties of multifold fermions survive disorder.
In time-reversal symmetric systems like RhSi, 
we cannot use real space invariants like the local Chern marker~\cite{Bianco2011} or the Bott index~\cite{Loring_2010},
used for time-reversal breaking Weyl semimetals~\cite{Yang2019, Zhang2021}.
Instead, we resort to the recently introduced spectral localizer~\cite{Schulz-Baldes2021, Schulz-Baldes2022, Cerjan2022}.

In three-dimensions, the spectral localizer is defined as~\cite{Schulz-Baldes2021, Cerjan2022}
\begin{equation}\label{eq:spectloc}
\mathcal{L} (\mathbf{r},E) = \kappa \sum_{j=1}^{3}  \gamma_j  (X_j-x_j \mathbb{I})  + \gamma_{4} (H - E \mathbb{I}),
\end{equation}
where $X_j$ are position operators corresponding to the Hamiltonian $H$, and the matrices $\gamma_j$ form a Clifford representation 
$\left\lbrace\gamma_j,\gamma_i\right\rbrace = 2\delta_{ij}$.  
We choose $\gamma_j = \tau_z \sigma_j$ for $j=1,2,3$ and $\gamma_{4}= \tau_x \sigma_0$, where $\sigma_j$ and $\tau_j$ are Pauli matrices. 
The coefficient $\kappa$ fixes the units and relative weights 
between $X_j$ and $H$~\cite{Cerjan2022,Loring2019}, see also the SM~\cite{SM} for a discussion concerning different values of $\kappa$.
While the spectral localizer can be evaluated at any position 
$\mathbf{r}$ and energy $E$, here we choose $\mathbf{r}$ 
to be at the center of our system ($\mathbf{r}=\mathbf{0} \rightarrow x_j =0 \; \forall j$) 
in order to probe the bulk properties at $E=0$.
In the following, we abbreviate $\mathcal{L} (\mathbf{0} ,0)$ with $\mathcal{L}_0$. 

The spectrum of $\mathcal{L}_0$ 
consists of pairs $(\epsilon, -\epsilon)$ because $\mathcal{L}_0$ 
obeys chiral symmetry $\mathcal{C} = \tau_y \sigma_0 \mathbb{I}$. 
In the crystalline limit, the parameter regime (1) yields 
four states pinned at $\epsilon=0$, 
that are separated from the remaining states by a gap of order $\sqrt{\kappa}$. 
Using a semi-classical analysis of the operator $\mathcal{L}_0^2$~\cite{Schulz-Baldes2022,SM},
it is possible to show analytically that each Weyl node contributes exactly one zero mode~\cite{Schulz-Baldes2022}.
We have generalized such an analysis~\cite{SM} for the case of a system with threefold 
and double-Weyl fermions, i.e., the a-RhSi regime, predicting four midgap modes as well~\cite{SM}.
The number of midgap modes of $\mathcal{L}_0$ can be thus used to signal multifold fermions in both regimes, 
as trivial metals present different midgap mode counting~\cite{Franca2023}.

\begin{figure}[t]
\includegraphics[width=8.6cm]{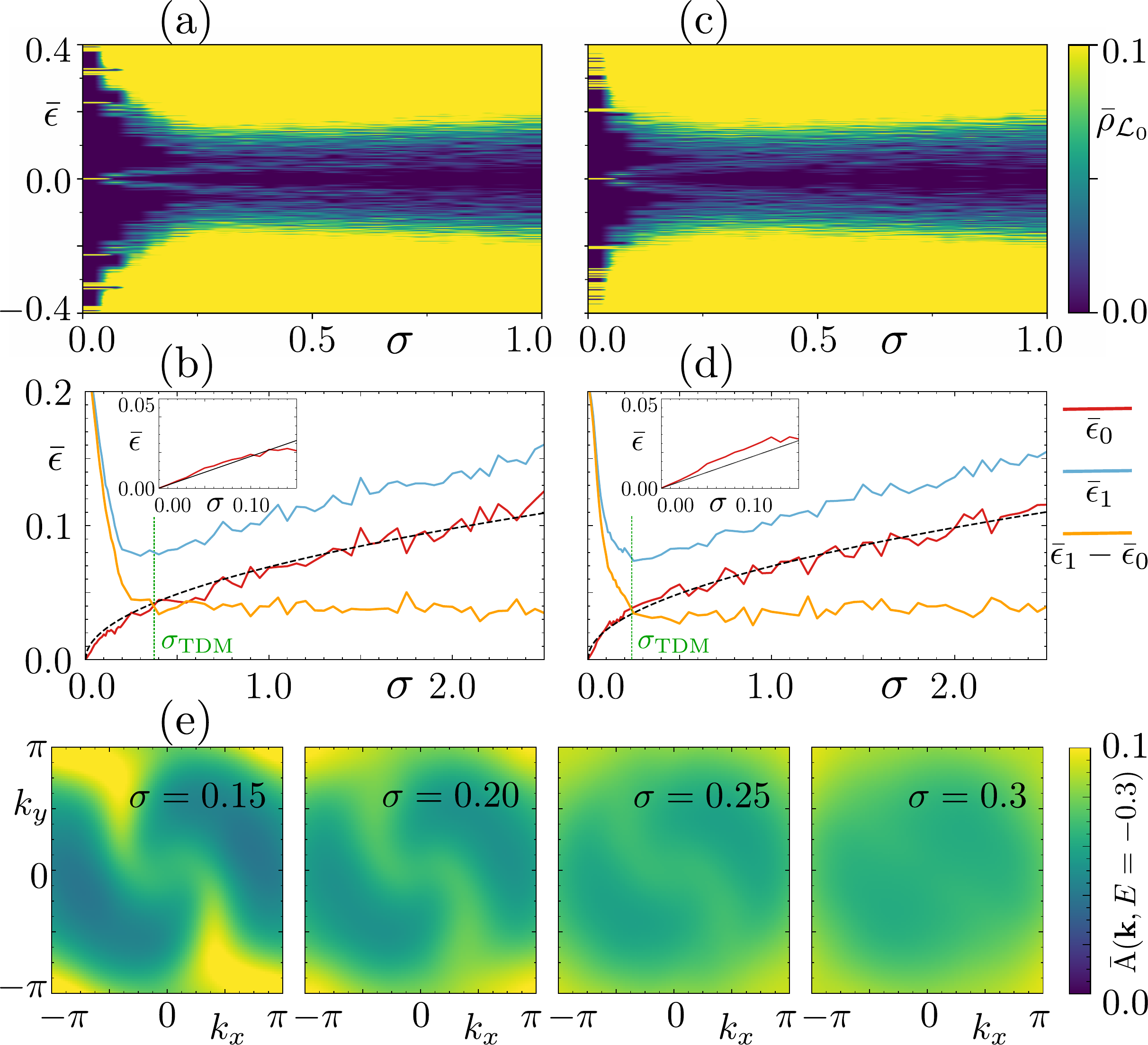}
\caption{
(a) and (c) show $\bar{\rho}_{\mathcal{L}_0}$ - the disorder averaged DOS of the operator $\mathcal{L}_0$ as a 
function of disorder strength $\sigma$ for regime (1) and the a-RhSi regime, respectively. 
In (b) and (d) are plotted disorder averaged energies of the midgap and first-excited states
$\bar{\epsilon}_0, \bar{\epsilon}_1$ and their difference $\bar{\epsilon}_1- \bar{\epsilon}_0$ as a function of $\sigma$ for the two parameter regimes.
Here, dashed green lines represent the topological phase transition point $\sigma_{\rm TDM}$ at which $ \bar{\epsilon}_1 \approx 2  \bar{\epsilon}_0$.
The dashed black line represents a fit $\bar{\epsilon}_0 = a \sqrt{\sigma}$, 
where $a = 0.0692$ in  regime (1) and $a= 0.0696$ for regime (2). 
The insets of (b) and (d) show how well $\bar{\epsilon}_0$ 
matches with the predicted form $\kappa^{0.75} \sigma$~\cite{Schulz-Baldes2022} 
(gray line) in case of small disorder strengths. 
For (a-d), we consider $\kappa = 0.1$ and $\rm{N_{dis} }= 25$.
In (e) are plotted disorder averaged ($\rm{N_{dis} }= 10$) momentum-resolved spectral functions $\bar{\rm A}(\mathbf{k}, E=-0.3)$ for different disorder strengths $\sigma$. }
\label{fig:fig4}
\end{figure}

To study how $\mathcal{L}_0$ changes with disorder, 
we focus on its DOS $\rho_{\mathcal{L}_0}$,
calculated using the KPM~\cite{Weisse2006} with an energy resolution $\Delta \epsilon = 5\times 10^{-4}$ ($N_C \sim 6000$).
The system size is $L=12$ and we consider $\rm N_{dis} = 25$ disorder realizations.
Fig.~\ref{fig:fig4}a shows the disorder averaged DOS 
$\bar{\rho}_{\mathcal{L}_0}$, as a function of disorder strength $\sigma$ for regime (1).
From Fig.~\ref{fig:fig4}a, we see that as $\sigma $ is increased, 
the four zero-energy states split into a pair of 
peaks that move away from $\epsilon=0$ in a symmetric fashion. 
In parallel, disorder reduces the spectral localizer gap and, at around $\sigma_{\rm TDM} = 0.35$, 
the energies of the midgap and first-excited states become comparable, 
indicating the transition into a trivial DM. 
The existence of the topological in-gap modes of $\mathcal{L}_0$ for 
$\sigma<\sigma_{\rm TDM}$ defines the topological diffusive metal phase (see Fig.~\ref{fig:fig1}a).

The transition from a TDM to a trivial DM is also apparent by tracking, 
for every disorder realization $\lambda$, the peak positions $\epsilon_0^{\lambda}, \epsilon_1^{\lambda}$ 
corresponding to the midgap mode and the first-excited state, respectively.
In Fig.~\ref{fig:fig4}b, we plot disorder averaged 
energies $\bar{\epsilon}_{0,1} = \frac{1}{\rm{N_{dis}}} \sum_{\lambda = 1}^{\rm{N_{dis}}} \epsilon_{0,1}^{\lambda}$ 
as a function of $\sigma$.
We see that $\bar{\epsilon}_0$ and $\bar{\epsilon}_1$ approach each other for small disorder strengths, and without crossing each other they start to evolve together with stronger disorder indicating a topologically trivial system. 
Since a topological phase is preserved with disorder as long as the dimensionality of the operator's $\mathcal{L}_0$ nullspace is nonzero, it is natural to assume that the topological phase transition occurs when $\bar{\epsilon}_1 - \bar{\epsilon}_0 \approx \bar{\epsilon}_0 \rightarrow \bar{\epsilon}_1 \approx 2 \bar{\epsilon}_0 $.
For the parameter regime (1), this occurs for $\sigma_{\rm TDM} \approx 0.35$.
In addition, for small disorder strengths, see inset of Fig.~\ref{fig:fig4}b, 
we find that $\bar{\epsilon}_0 = \sigma \kappa^{3/4}$ 
consistent with the analytical prediction concerning weakly disordered 
Weyl semimetals~\cite{Schulz-Baldes2022}.
Moreover, we find that $\bar{\epsilon}_0 $ can be fitted with a function $ a \sqrt{\sigma} $ ($a \approx 0.07$) in the entire disorder range.

In Figs.~\ref{fig:fig4}c,d we show results for the a-RhSi regime. 
Even though the system supports a threefold fermion in the crystalline limit, 
$\bar{\rho}_{\mathcal{L}_0}$ behaves similarly to regime (1) with two double-Weyl fermions.
From Fig.~\ref{fig:fig4}d, we see that $\sigma_{\rm TDM} \approx 0.25$.
Furthermore, we recover $\bar{\epsilon}_0 = \sigma \kappa^{3/4}$ behavior in the limit of small disorder, as well as 
$\bar{\epsilon}_0 \propto \sqrt{\sigma}$ for the entire disorder range.

To confirm that the spectral localizer correctly captures the topological phase transition, we study how the Fermi arcs at the top surface of a system in a-RhSi regime evolve with disorder.
These arcs can be seen with the disorder averaged momentum-resolved spectral function $\bar{\rm A}(\mathbf{k} , E = -0.3)$ that can be measured in angle-resolved photoemission experiments~\cite{Marsal2023}. 
The plots of $\bar{\rm A}(\mathbf{k} , E = -0.3)$ in Fig.~\ref{fig:fig4}e for disorder strengths $\sigma = 0.15, 0.2,0.25,0.3$ indicate that the Fermi arcs disappear for $\sigma > 0.25$, in agreement with the prediction of $\mathcal{L}_0$.
For more details, see the SM~\cite{SM}.

We find that both regimes behave similarly as long the hopping amplitude $v_p$ 
is the largest energy scale.
This condition ensures a sizable difference between the amplitudes of direction 
dependent nearest-neighbor hoppings $(v_1 \pm v_p)/4$.
This condition is met by the RhSi parameters but not for those of CoSi, where the parameter $v_1$ is more than three times larger 
than the parameter $v_p$. 
As a result, the topological properties of a-CoSi are expected to be less robust compared to a-RhSi~\cite{SM}.

\textit{\textcolor{blue}{Conclusion}} --
We have shown that a topological type of diffusive metal can
exist in transition metal monosilicides in the
presence of structural, potential and hopping disorder.
Characterizing this novel phase required us to extended the 
recently discovered spectral localizer $\mathcal{L}$ to 
accommodate multifold fermions.
The spectral localizer can be used to signal TDMs in any symmetry class, 
including those for which other real-space methods yield trivial results or are ill-defined.
Our analysis highlights a-RhSi as a more robust platform than a-CoSi to realize
the topological diffusive metal 
due to a larger anisotropy between nearest neighbor hoppings.
Looking forward, it is worth studying
whether such stability permeates to physical properties 
such as the photogalvanic effect or negative magneto-resistance.

\textit{\textcolor{blue}{Acknowledgments}} --
We thank Q. Marsal, H.~Schulz-Baldes, J. Wilson and D. Carpentier for discussions. 
A.G.G. and S. F. acknowledge financial support from the European Union Horizon 2020 
research and innovation program under grant agreement No. 829044 (SCHINES).
A. G. G. is also supported by the European Research Council (ERC) 
Consolidator grant under grant agreement No. 101042707 (TOPOMORPH).
The Kwant code~\cite{Groth2014} used to generate our results is available at~\cite{zenodo}.

\bibliography{NHr}

\clearpage
\newpage

\appendix

\title{Supplemental Material to:  Topological diffusive metal in amorphous transition metal monosilicides }

\maketitle

\tableofcontents

\setcounter{secnumdepth}{5}
\renewcommand{\theparagraph}{\bf \thesubsubsection.\arabic{paragraph}}

\renewcommand{\thefigure}{S\arabic{figure}}
\setcounter{figure}{0}

\section{The Bloch Hamiltonian}\label{sec:sec1}

We begin this section with classifying, by separation distance, all neighboring orbitals of a crystalline system within radius $d_c = 1.5$, where $a$ is the lattice constant. 
For each of these distances, we provide a Bloch Hamiltonian for a hopping term between two orbitals. 
To show that these Bloch Hamiltonians correctly describe the bulk of a finite crystal, 
we compare their band structure with the momentum-resolved spectral function $A(\mathbf{k},E)$.

The radial distribution function $g(d)$ gives us the probability of finding two sites separated by distance $d$.
It is defined as 
\begin{equation}
g(d) = \frac{\delta n_{d}}{\delta V_{d} \eta},
\end{equation}
where $\delta n_{d}$ represents the number of sites within a spherical shell of thickness $\delta d$ 
and volume $\delta V_{d} = 4\pi d^2 \delta d$, and $\eta$ is the bulk density of sites. 
We measure all distances in units of lattice constant $a$.

\begin{figure}[tb]
\includegraphics[width=8.6cm]{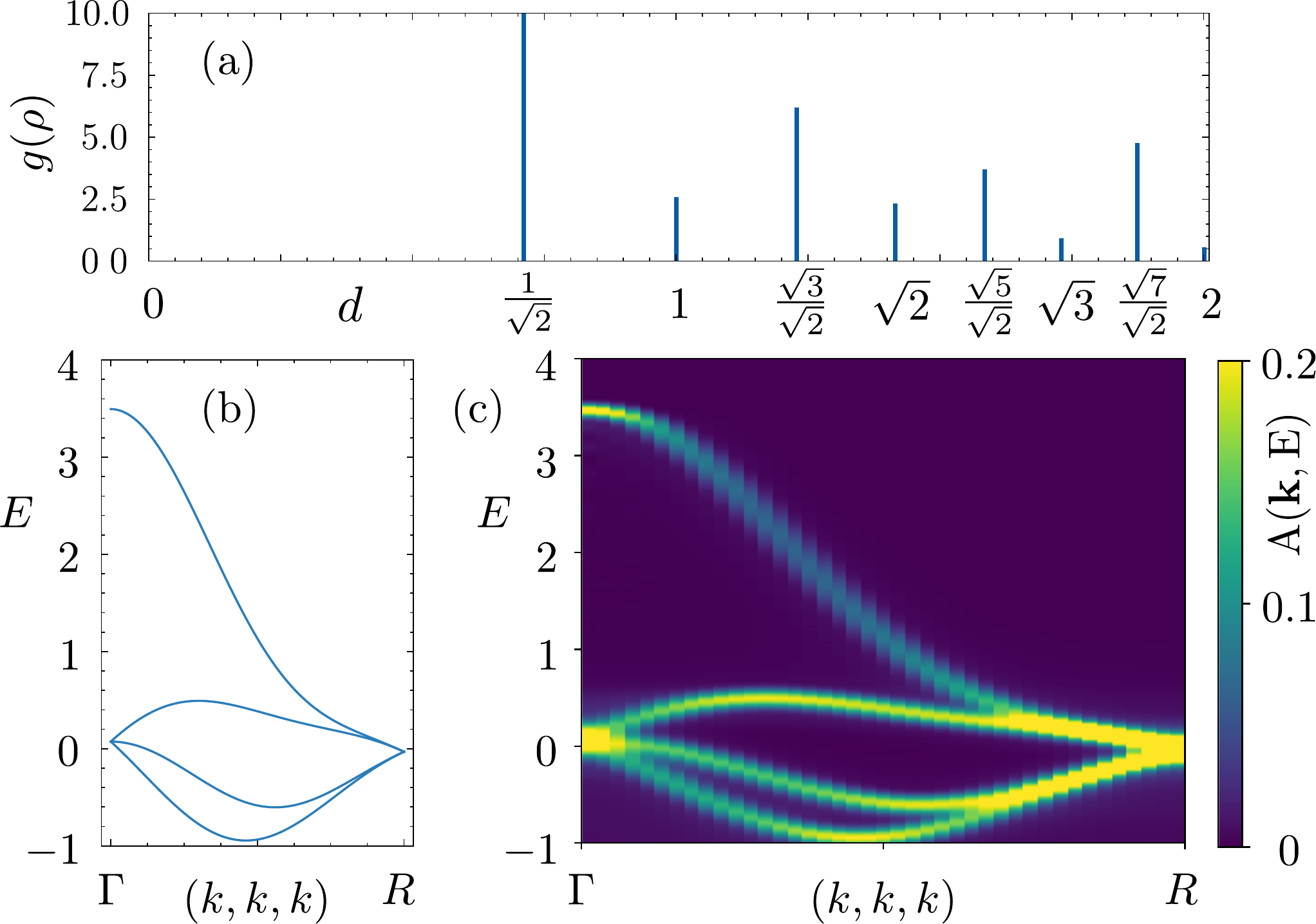}
\caption{Panel (a) shows the radial distribution function $g(d)$ as a function of distance $d$ for the crystalline transition metal monosilicide.
Panel (b) shows the band structure for the crystalline RhSi ($v_ 1 = 0.55, v_2 = 0.16$ and $v_p = -0.762$) 
and a cutoff distance $d_c = 1.5$ along path $\Gamma-R$ in the Brillouin zone, 
calculated using the Bloch Hamiltonian $\mathcal{H}$ with up to fourth-nearest neighbours.
Panel (c) shows $A(\mathbf{k},E)$ along the same path in BZ, and for the same set of parameters, calculated using the 
real-space Hamiltonian $H$.}
\label{fig:figsm0}
\end{figure}

In Fig.~\ref{fig:figsm0}a, we show $g(d)$ for a crystalline transition metal monosilicide. 
We see that the first nearest neighboring orbitals are located at a distance $d =1/\sqrt{2}$.
As discussed in the main text, the hopping between these orbitals are angle dependent as described by the Hamiltonian~\cite{Chang2017} 
\begin{widetext}
\begin{equation} \label{eq:H1}
\begin{split}
\mathcal{H}_1 =   & v_1 [\gamma_x \delta_0 \cos{\frac{k_x}{2}} \cos{\frac{k_y}{2}}+ \gamma_x \delta_x \cos{\frac{k_y}{2}} \cos{\frac{k_z}{2}}+
\gamma_0 \delta_x \cos{\frac{k_z}{2}} \cos{\frac{k_x}{2}}] + \\
&  v_p [\gamma_y \delta_z \cos{\frac{k_x}{2}} \sin{\frac{k_y}{2}}+ \gamma_y \delta_x \cos{\frac{k_y}{2}} \sin{\frac{k_z}{2}}+
\gamma_0 \delta_y \cos{\frac{k_z}{2}} \sin{\frac{k_x}{2}}],
\end{split}
\end{equation}
\end{widetext}
that is written in the basis $\Psi = (c_A, c_B, c_C, c_D)$, where $c_{\alpha}$ represents the annihilation operator of a particle at orbital $\alpha = A, B, C, D$.

The second nearest-neighbor hopping connects orbitals of the same kind, separated by a distance $d = 1$.
These hoppings are captured by the Bloch Hamiltonian~\cite{Chang2017}
\begin{equation} \label{eq:H2}
\begin{split}
\mathcal{H}_2 =  v_2 [\cos{k_x} + \cos{k_y} + \cos{k_z}] \gamma_0 \delta_0. 
\end{split}
\end{equation}

{Since we allow the cut-off radius to vary, our model may involve longer range hoppings.
Thus we extend the model of Ref.~\cite{Chang2017} to incorporate longer-range hoppings, i.e, beyond second-nearest neighbors.}

All the longer-range hoppings can also be split into inter-orbital and intra-orbital hoppings,
and we determine their amplitudes following Eq.~(2) of the main text.
For completeness, we repeat it here
\begin{equation}~\label{eq:hopdef1_sm}
\tilde{v}_{\alpha} = \tilde{v}_{\alpha} (d) \tilde{v}_{\alpha} (\theta, \phi) \exp[1-\frac{d}{d_{\alpha} ^0}] \Theta(d_c-d),
\end{equation}
where $d, \theta$ and $\phi$ are the relative spherical coordinates between the positions of the two sites 
involved in the hopping.
Their radial and angular dependencies are given in Table~\ref{table:1}.

\begin{table}[h!]
\centering
\begin{tabular}{|c c c c|} 
 \hline
$\alpha$ & $\tilde{v}_{\alpha}  (d)$ & $ \tilde{v}_{\alpha} (\theta, \phi)$  & orbitals involved \\ [0.5ex] 
 \hline\hline
 $1$ & $ v_1/\sqrt{2} d$ & $1$ & $A \leftrightarrow B/C/D; B \leftrightarrow C/D, C \leftrightarrow D$ \\ 
 $2$ & $v_2/d$ &$1$ & $A \leftrightarrow A; B \leftrightarrow B; C \leftrightarrow C; D \leftrightarrow D$  \\
 $p$ & $v_p/d$ & $\sin{\theta} \cos{\phi}$ & $B \rightarrow A/ D \rightarrow C$ \\
 $p$ & $v_p/d$ & $\sin{\theta} \sin{\phi}$ & $C \rightarrow A/ B \rightarrow D$ \\
 $p$ & $v_p/d$ & $\cos{\theta}$ & $D \rightarrow A/ C \rightarrow B$ \\ [1ex] 
 \hline
\end{tabular}
\caption{Radial and angular amplitudes of hoppings in transition metal monosilicides. 
For hoppings $v_p$, we need to explicitly state the starting and target orbitals in order to fix the signs of $\tilde{v}_{\alpha} (\theta, \phi)$. 
For example $B \rightarrow A$ implies that the particle hops from a site of orbital type B to a site of orbital type A with an angular hopping amplitude $\sin{\theta} \cos{\phi}$. 
By Hermitian conjugation, this also sets the hopping amplitude from a site of orbital type A to a site of orbital type B.}
\label{table:1}
\end{table}

Third nearest-neighbor hoppings connect orbitals at a distance $d= \sqrt{3/2}$.
Following Eq.~\eqref{eq:hopdef1_sm} and Table~\ref{table:1}, the hopping amplitudes for a crystalline lattice read $\frac{v_1}{\sqrt{3}} \exp[1-\sqrt{3}]$ and $\frac{v_p}{3} \exp[1-\sqrt{3}]$.
The corresponding Bloch Hamiltonian reads
\begin{widetext}
\begin{equation} \label{eq:H3}
\begin{split}
\mathcal{H}_{3} =   &  \frac{ v_1}{\sqrt{3}}   e^{1-\sqrt{3}}  
 [\gamma_x \delta_0 \cos{\frac{k_x}{2}} \cos{\frac{k_y}{2}}2 \cos{k_z}+ 
 \gamma_x \delta_x 2 \cos{k_x}\cos{\frac{k_y}{2}} \cos{\frac{k_z}{2}}+
\gamma_0 \delta_x \cos{\frac{k_z}{2}} 2 \cos{k_y}\cos{\frac{k_x}{2}}] + \\
& \frac{v_p}{3} e^{1-\sqrt{3}}  
 [\gamma_y \delta_z \cos{\frac{k_x}{2}} \sin{\frac{k_y}{2}}2\cos{k_z}+
  \gamma_y \delta_x 2\cos{k_x} \cos{\frac{k_y}{2}} \sin{\frac{k_z}{2}}+
\gamma_0 \delta_y \cos{\frac{k_z}{2}} 2\cos{k_y} \sin{\frac{k_x}{2}}].
\end{split}
\end{equation}
\end{widetext}

Orbitals of the same type at a distance $d = \sqrt{2}$ are fourth nearest-neighbors,
and are related by hopping amplitudes $\frac{v_2}{\sqrt{2}} \exp[1-\sqrt{2}]$, following Eq.~\eqref{eq:hopdef1_sm} and Table~\ref{table:1}. 
The Bloch Hamiltonian reads
\begin{widetext}
\begin{equation} \label{eq:H4}
\begin{split}
\mathcal{H}_4 = 2 \frac{v_2}{\sqrt{2}} \exp[1-\sqrt{2}] [\cos{k_x}\cos{k_y} + \cos{k_y}\cos{k_z} + \cos{k_z}\cos{k_x} ] \gamma_0 \delta_0. 
\end{split}
\end{equation}
\end{widetext}

In our simulations, we choose a cutoff at $d_c= 1.5$ such that the Bloch Hamiltonian 
$\mathcal{H} = \mathcal{H}_1+\mathcal{H}_2+\mathcal{H}_3+\mathcal{H}_4$ describes bulk properties of a crystalline transition metal monosilicide. 
In the following, we focus on RhSi in parameter regime (2). 
Its band structure is shown in Fig.~\ref{fig:figsm0}b.
Compared to Ref.~\cite{Chang2017}, where $\mathcal{H} = \mathcal{H}_1+\mathcal{H}_2$, 
we see that when longer-range hoppings are added the threefold fermion 
at $\Gamma$ is closer in energy to the double-Weyl fermion at $R$.

To verify that this band structure accurately describes the bulk of the constructed crystalline system, 
we calculate the momentum-resolved spectral function $A(\mathbf{k},E)$ for a finite system.
The spectral function is calculated by projecting the real space spectral function onto a plane-wave basis~\cite{Marsal2023, Corbae2023a}
\begin{equation}\label{eq:spect_func}
A(\mathbf{k}, E) = \sum_{o = A,B,C,D}  \braket{\mathbf{k}_o \big| \delta(H-E)}{\mathbf{k}_o},
\end{equation} 
where $\rm H = H_1+H_2+H_3+H_4$ and $\rm H_1,H_2, H_3, H_4$ are real space tight-binding Hamiltonians deduced from Eqs.~\eqref{eq:H1},~\eqref{eq:H2},~\eqref{eq:H3} and~\eqref{eq:H4}, respectively.
Moreover $\ket{\mathbf{k}_o} = \frac{1}{\sqrt{V}}\exp[i \mathbf{k} \mathbf{r}_{n,o}]\ket{n,o}$ is a vector of $V = 4 L^3$ entries corresponding to positions $ \mathbf{r}_n$ of sites within the system. 
These entries are nonzero only for those sites belonging to an orbital type $o = A,B,C,D$.
The spectral function can be measured in angle-resolved photoemission (ARPES) experiments, where momentum $\mathbf{k}$ represents the momentum of the photo-emitted electron~\cite{Marsal2023, Corbae2023a}. 

For a system with linear size $L = 20$ and under periodic boundary conditions (PBCs), we plot $A(\mathbf{k},E)$ in Fig.~\ref{fig:figsm0}c.
We observe that the spectral function calculated with $H$ captures well the energies of threefold and double-Weyl fermions in Fig.~\ref{fig:figsm0}b, calculated with $\mathcal{H}$, as well as the bandwidth of each band.


\section{Rare states}
In this section, we demonstrate the existence of rare states~\cite{Pixley2016} in a-RhSi in parameter regime (1). 

In this parameter regime, the density of states of the crystalline system vanishes at $E=0$.
Once disorder strength $\sigma$ is non-zero, statistically rare regions, with exceptionally strong disorder strength, are possible. 
Because $\rho(E=0)=0$, these regions may trap states at low energies that we call rare states. 
To confirm the existence of these rare states for our amorphous systems, we follow the procedures outlined in Ref~\cite{Pixley2016}.

In finite systems, the energy scales of rare states and the double-Weyl fermions are comparable 
thus complicating the detection of rare states. 
For this reason, we study the spectra of infinite systems obtained by imposing twisted boundary conditions.
These boundary conditions assume that hoppings between outermost orbitals belonging to different surfaces are $v_{\alpha} \exp[i \theta_t]$ $(\alpha = 1,2,p)$, where $\theta$ is called the twist angle. 
For $\theta_t= 0/\pi$, we say the system in under periodic/anti-periodic boundary conditions. 
As explained in Ref~\cite{Pixley2016}, combining appropriate boundary conditions with different system sizes 
we can maximally separate the energy scales of double-Weyl fermions and disorder induced rare-states.

\begin{figure}[tb]
\includegraphics[width=8.6cm]{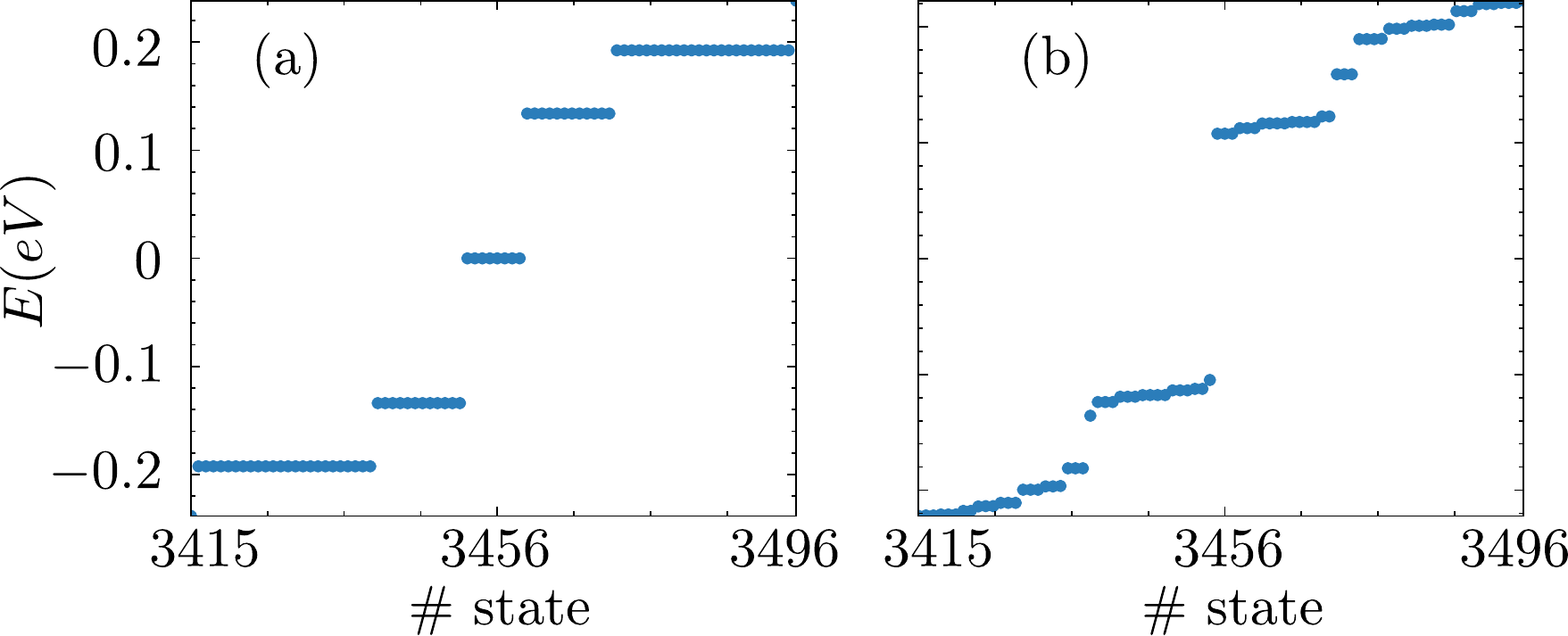}
\caption{The low-energy spectrum of a crystalline system in parameter regime (1) with double-Weyl fermions pinned at $E=0$ under (a) periodic boundary conditions and (b) anti-periodic boundary conditions. 
Both types of boundary conditions are applied in all three directions.
Here, we consider a cubic system with $12^3$ unit cells.}
\label{fig:figsm1}
\end{figure}

In the following, we consider a crystalline system with an even number of unit cells $L$ in all directions. 
In the limit $\theta_t = 0$, this system has eight gapless states, see Fig.~\ref{fig:figsm1}a.
These states reflect the existence of two zero-energy double-Weyl fermions at the $\Gamma$ and $R$ points of the Brillouin zone. 
Applying anti-periodic boundary conditions (APBCs) in all directions shifts these states away from $E=0$, as shown in Fig.~\ref{fig:figsm1}b.
This creates an energy gap that fills up with rare states once $\sigma \neq 0$.  

\begin{figure}[tb]
\includegraphics[width=8.6cm]{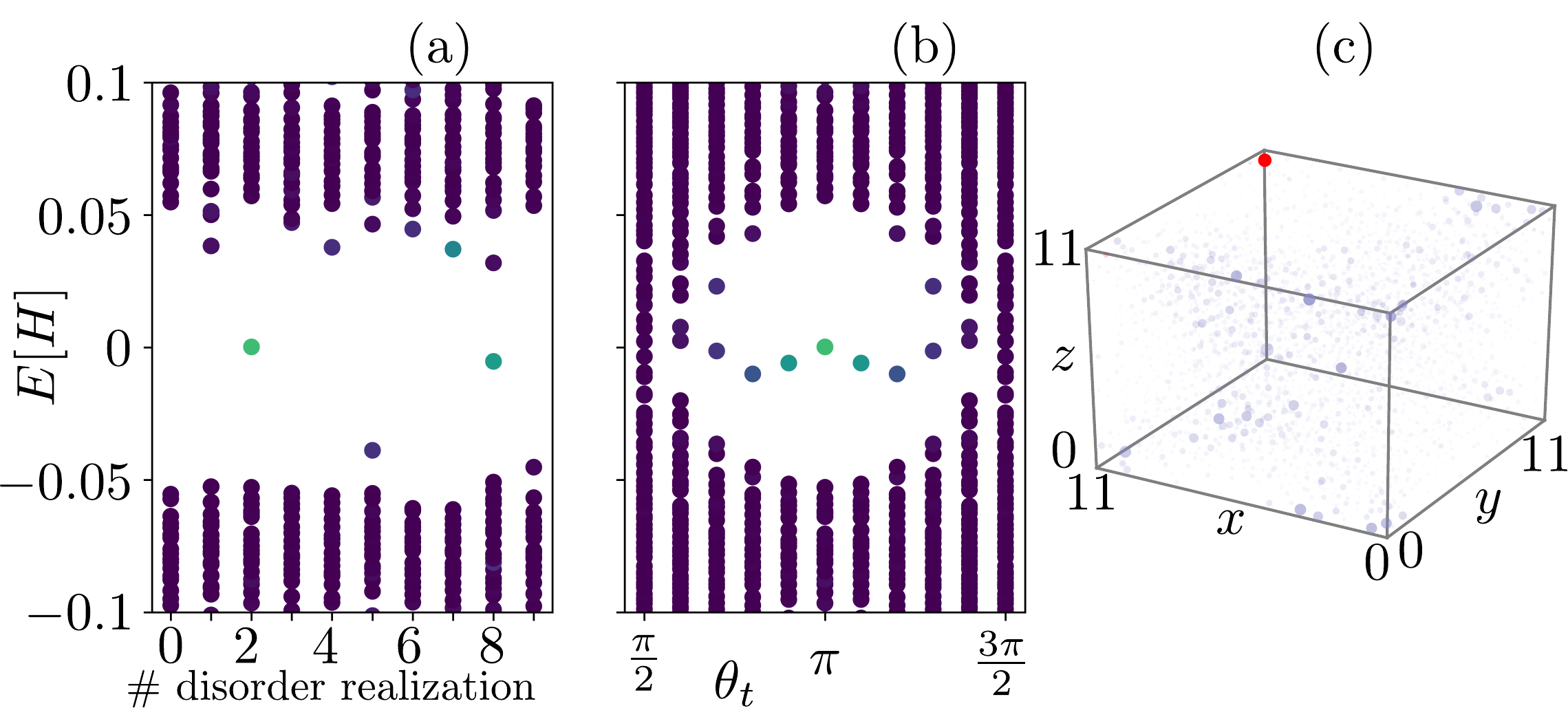}
\caption{The low-energy spectrum of an amorphous system in parameter regime (1) under (a) anti-periodic boundary conditions and (b) twisted boundary conditions.
Panel (c) shows probability density distributions several states of the amorphous system created in disorder realization $\lambda = 2$, which spectrum is shown in panel (a).
Here, red color represents the low-energy state near $E=0$, and blue color represents joint probability density distributions of three excited states lowest in energy. 
We assume a cubic system with $12^3$ unit cells.}
\label{fig:figsm2}
\end{figure}

We now study the amorphous system in regime (1), with $12^3$ unit cells and with APBCs in all directions. 
In Fig.~\ref{fig:figsm2}a, we plot the Hamiltonian spectrum as a function of different disorder realizations for $\sigma=0.07$. 
We color all eigenstates $m$ of this spectrum according to their inverse participation ratio (IPR) 
$\rm IPR_m = \sum_{\mathbf{r}_n}  |\Psi_m (\mathbf{r}_n)|^4$, such that blue and green color indicates smaller and larger IPR, respectively. 
States that are shifted away from $E=0$ have low IPR, indicating they are perturbatively dressed Weyl states~\cite{Pixley2021}.
For disorder realizations $\lambda = 2, 8$, we observe two low-energy states with increased localization compared to the higher-energy states. 

In the following, we focus on disorder realization $\lambda = 2$. 
The low-energy spectrum of this system as a function of twist angle $\theta_t$ is shown in Fig.~\ref{fig:figsm2}b. 
We see that a low-energy state disperses weakly with $\theta_t$ compared to higher-energy states, suggesting it is a localized state~\cite{Pixley2015}.
Fig.~\ref{fig:figsm2}c shows its probability density distribution (red color) at $\theta_t = \pi$ along with the joint probability distribution of three excited states lowest in energy (blue color). 
We see that the low-energy state is localized in a small region, while excited states spread over large portions of the system.

\section{Semi-classical analysis of the spectral localizer}

In this section, we analytically derive the low-energy spectrum of the spectral localizer for crystalline transition metal monosilicides.
This can be done using a semi-classical analysis that maps the problem of finding zero modes of the operator $\mathcal{L}$ to a well-known problem of solving for the spectrum of harmonic oscillators~\cite{Schulz-Baldes2021, Schulz-Baldes2022, Franca2023}.
As in ~\cite{Schulz-Baldes2021, Schulz-Baldes2022}, semi-classical refers to the fact 
that we ignore tunneling between between harmonic oscillator wells, 
ensured by choosing a small value of $\kappa$, as defined next.
We start with the spectral localizer 
introduced in Eq.~(4) of the main text in the limit $\mathbf{r} = 0$ and $E = 0$ such that we probe bulk properties near the topological crossings.
We rewrite it here in the following form
\begin{equation}\label{eq:spectloc}
\mathcal{L} = 
\begin{pmatrix}
\kappa D & \sigma_0 H \\
\sigma_0 H & -\kappa D
\end{pmatrix},
\end{equation}
where we have omitted the subscript $0$ in $\mathcal{L}_0$, in order to simplify the notation in this section.  
We define $D = \sum_j \sigma_j X_j$, with $X_j$ and $H$ being the real-space position and Hamiltonian operators, respectively.
Next, we perform a Fourier transform $\mathcal{F}$ on $\mathcal{L}$.
Using $\mathcal{F}[D] = - i \sum_{j=1}^d \sigma_j \partial_{k_j}$ and $\mathcal{H} = \mathcal{F}[H]$, we obtain
\begin{equation}\label{eq:spectloc_kspace}
\mathcal{L}_k = 
\begin{pmatrix}
- i \kappa \sum_j \sigma_j \partial_{k_j}  & \sigma_0 \mathcal{H} \\
\sigma_0\mathcal{H} & i \kappa \sum_j \sigma_j \partial_{k_j}
\end{pmatrix}.
\end{equation}
We study the spectral properties of the operator
 \begin{equation}\label{eq:spectloc_squared}
\mathcal{L}_k^2 = 
\begin{pmatrix}
\sigma_0 (\mathcal{H}^2 - \kappa^2 \nabla_k^2) &- i \kappa  \sum_{j=1}^d \sigma_j \partial_{k_j} \mathcal{H}\\
i \kappa  \sum_{j=1}^d \sigma_j \partial_{k_j} \mathcal{H} & \sigma_0 (\mathcal{H}^2 - \kappa^2 \nabla_k^2)
\end{pmatrix}
\end{equation}
that is, by construction, non-negative and, 
as will be shown shortly, has harmonic potential wells at Weyl points~\cite{Schulz-Baldes2021}.

\medskip

\begin{figure*}[tb!]
\includegraphics[width=\textwidth]{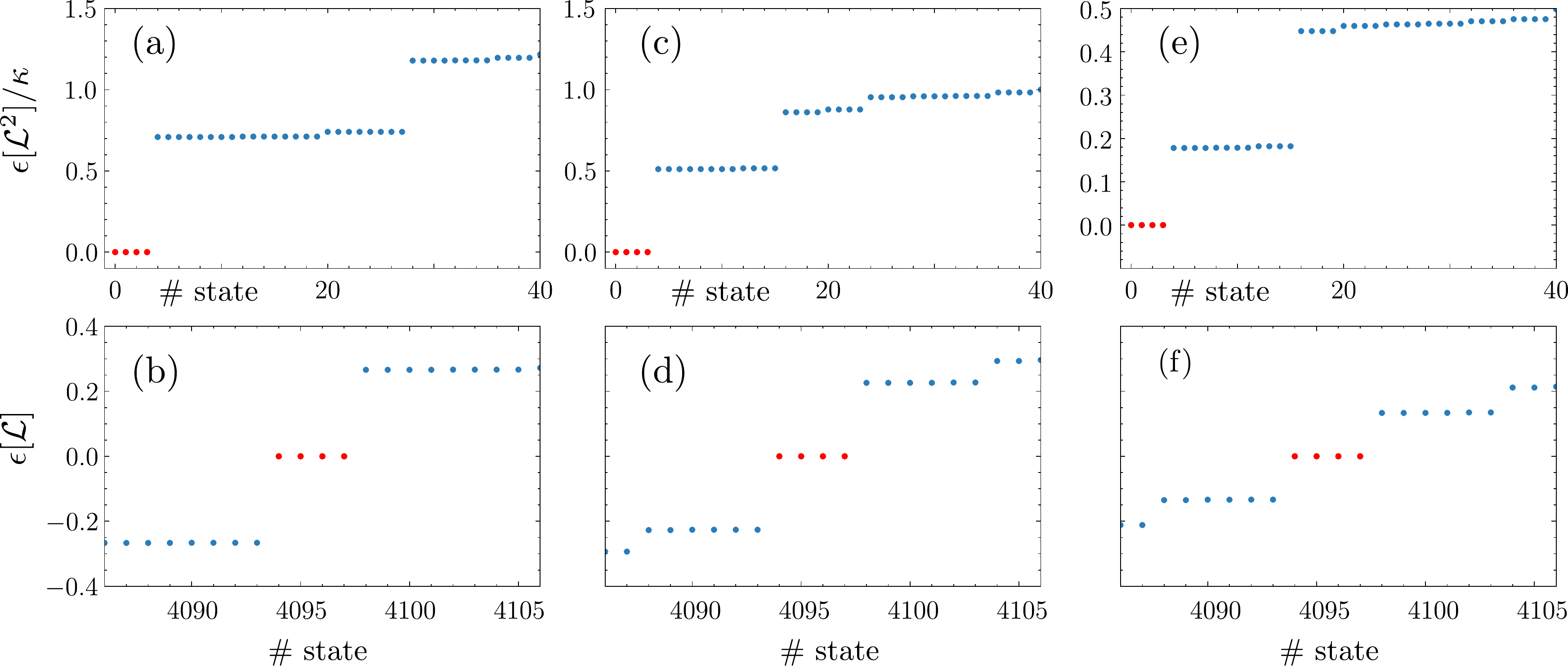}
\caption{
Panels (a, c, e) and (b, d, f) show the spectra of the operators $\mathcal{L}^2$ and $\mathcal{L}$, respectively.
These operators are calculated for $\kappa = 0.1$ and a crystalline system of linear size $L = 8$.  
Panels (a, b) and (c, d) consider parameter regime (1) and has $d_c  1.01$ and $d_c = 1.5$, respectively. 
Panels (e,f), consider parameter regime (2) with $d_c= 1.01$.
We use hopping amplitudes $v_p = -0.762$ and $v_2 - 0.16$.
Red and blue color represents zero modes and excited states, respectively. 
}
\label{fig:figsm7}
\end{figure*}

\subsection{Parameter regime (1)}

To build intuition we focus on crystalline systems in parameter regime (1) with nonzero hoppings $v_p$. 
To simplify our analysis, we consider the case with cutoff radius $d_c= 1.01$ such that only nearest-neighbors are related by nonzero hoppings. 
From Eq.~\eqref{eq:H1}, we get  
\begin{widetext}
\begin{equation} \label{eq:ham_regime1}
\mathcal{H}=  v_p [\gamma_y \delta_z \cos{\frac{k_x}{2}} \sin{\frac{k_y}{2}}+ \gamma_y \delta_x \cos{\frac{k_y}{2}} \sin{\frac{k_z}{2}}+
\gamma_0 \delta_y \cos{\frac{k_z}{2}} \sin{\frac{k_x}{2}}].
\end{equation}
\end{widetext}

We are interested in the low-energy counterpart of this Hamiltonian in order to study the low-energy spectrum of $\mathcal{L}_k^2 $.
To probe a double-Weyl fermion at the $\Gamma$ point, we expand $\mathcal{H}$ close to $\mathbf{k}= 0$ to first order in $\mathbf{k}$ and obtain
\begin{equation}\label{eq:lowEham_regime1}
\mathcal{H} =  v_p [\gamma_0 \delta_y \frac{k_x}{2} + \gamma_y \delta_z \frac{k_y}{2}+ \gamma_y \delta_x \frac{k_z}{2}].
\end{equation}
We can bring this Hamiltonian into a block-diagonal form characteristic for double-Weyl fermions~\cite{Bradlyn2016}, 
by performing a unitary transformation $U \mathcal{H} U^{\dagger}$ with $U = \frac{1}{\sqrt{2}}(\gamma_0-i \gamma_x) \delta_0$ and redefining momenta as $k_x \rightarrow k_y \rightarrow k_z \rightarrow k_x$.
This results in
\begin{equation}
\mathcal{H} = \frac{v_p}{2} [ \gamma_z \delta_x k_x+ \gamma_0 \delta_y k_y + \gamma_z \delta_z k_z ].
\end{equation}
From here, we readily obtain the operator $\mathcal{L}_k^2 $ as
\begin{widetext}
 \begin{equation}\label{eq:spectloc_dWFs}
\mathcal{L}_k^2 = 
\begin{pmatrix}
\sigma_0\gamma_0 \delta_0 [\frac{v_p^2}{4} (k_x^2+k_y^2+k_z^2) - \kappa^2 \nabla_k^2]
&- i \frac{v_p \kappa}{2} (\sigma_x  \gamma_z \delta_x+ \sigma_y  \gamma_0 \delta_y + \sigma_z \gamma_z \delta_z) \\
i \frac{v_p \kappa}{2} (\sigma_x  \gamma_z \delta_x+ \sigma_y  \gamma_0 \delta_y + \sigma_z \gamma_z \delta_z)
 & \sigma_0 \gamma_0 \delta_0 [\frac{v_p^2}{4} (k_x^2+k_y^2+k_z^2)  - \kappa^2 \nabla_k^2]
\end{pmatrix}.
\end{equation}
\end{widetext}
We observe that the diagonal elements $ [ - \kappa^2 \nabla_k^2 + \frac{v_p^2}{4} (k_x^2+k_y^2+k_z^2)]$ describe the equation of a 3D quantum harmonic oscillator in momentum space upon identifying $\hbar =\kappa , m = \frac{1}{2}$ and $\omega = v_p$.
The eigenvalues of this harmonic oscillator are $\epsilon^{\rm d}_n = \hbar \omega (n+\frac{3}{2}) = \kappa v_p (n+\frac{3}{2})$ where $n \in \mathbb{N}$. 
The lowest energy $\epsilon^{\rm d}_0 = \frac{3  v_p \kappa}{2}$ is obtained for $n=0$. 
Note that these eigenvalues are sixteen-fold degenerate because they originate from the diagonal part of $\mathcal{L}_k^2$.

The eigenvalues of the off-diagonal part can be obtained by its diagonalization.
The sixteen eigenvalues consists of doubly degenerate $\pm \frac{3 v_p \kappa}{2}$ and sixfold degenerate $\pm \frac{ v_p \kappa}{2}$. 
Thus, because diagonal and off-diagonal parts of operator $\mathcal{L}_k^2$ commute, its spectrum can be obtained by simply summing their respective eigenvalues, i.e., $\epsilon_n =\epsilon^{\rm d}_n + \epsilon^{\rm off-d}$. 
For $n=0$, we see that the doubly degenerate $\epsilon^{\rm off-d} = - \frac{3 v_p \kappa}{2}$ exactly cancels out the diagonal contribution $\epsilon^{\rm d}_0 = \frac{3  v_p \kappa}{2}$, resulting in two zero modes for the operator $\mathcal{L}_k^2$. 
Thus, we obtain that a single double-Weyl fermion at $\Gamma$ point yields two zero modes of operator $\mathcal{L}_k^2$, and via the inverse Fourier transformation, also for the operator $\mathcal{L}^2$. 
Similarly, the double-Weyl fermion at $R$ point also contributes two zero modes of the operators $\mathcal{L}_k^2$ and $\mathcal{L}$.
Hence, we can predict analytically that the operator $\mathcal{L}$ will have four zero modes for crystalline systems described by the Hamiltonian Eq.~\eqref{eq:ham_regime1}.

The presence of longer range hoppings, such as the third nearest neighbor hoppings does not alter this conclusion.
This is because such hopping {terms, given by Eq.~\eqref{eq:H3}, have the same matrix structure as the Hamiltonian Eq.~\eqref{eq:ham_regime1}.
Hence, when expanding up to linear order in $k$, the matrix structure remains the same as for the Hamiltonian Eq.~\eqref{eq:lowEham_regime1}, and the only difference is in the prefactors.
Thus, longer range hoppings with amplitudes proportional $v_p$ can only alter states higher in the spectrum of $\mathcal{L}$.}

We confirm this analytical prediction numerically for $v_p = -0.762$.
Figs.~\ref{fig:figsm7}a and b show the spectrum of operators $\mathcal{L}^2$ and $\mathcal{L}$, respectively for the system with linear size $L = 8$ and $\kappa=0.1$ in case of nonzero nearest-neighbor hoppings.
Furthermore, we obtain similar spectra once the third-nearest-neighbor hoppings are included, see Figs.~\ref{fig:figsm7}c, d.
We see that adding more hoppings preserves the existence of four degenerate zero modes but reduces the gap between them and states at higher energy.

\begin{widetext}

\subsection{Parameter regime (2)}

Next, we study how including hoppings with amplitudes $v_1, v_2$ affect the spectrum of the operator $\mathcal{L}$. 
For simplicity, we assume that only nearest-neighbor and second-nearest-neighbor hoppings are present. 
In parameter regime (2), the crystalline system hosts 
a threefold fermion at the $\Gamma$ point and a double-Weyl fermion at the $R$ point.
Since we derived the low-lying localizer spectrum for a double-Weyl in the previous section, here we focus 
on the $\Gamma$ point.
Close to $\Gamma$, the low-energy Hamiltonian reads

\begin{equation} \label{eq:Hk}
\mathcal{H} =    3 v_2 \gamma_0 \delta_0+ v_1 [\gamma_x \delta_0+ \gamma_x \delta_x +
\gamma_0 \delta_x ] + 
  \frac{v_p}{2} [\gamma_y \delta_z k_y + \gamma_y \delta_x k_z+
\gamma_0 \delta_y k_x].
\end{equation}

As before, we perform the unitary transformation on $\mathcal{H}$ with $U = \frac{1}{\sqrt{2}}(\gamma_0-i \gamma_x) \delta_0$ and
exchange momenta as $k_x \rightarrow k_y \rightarrow k_z \rightarrow k_x$.
This yields
\begin{equation} \label{eq:Hk}
\mathcal{H}=    3 v_2 \gamma_0 \delta_0+ v_1 [\gamma_x \delta_0+ \gamma_x \delta_x +
\gamma_0 \delta_x ] + 
  \frac{v_p}{2} [\gamma_z \delta_x k_x+ \gamma_0 \delta_y k_y + \gamma_z \delta_z k_z].
\end{equation}

After squaring this Hamiltonian, we perform a unitary transformation $U \mathcal{H}^2 U^{\dagger} = \tilde{\mathcal{H}}^2 $ with $U = \frac{1}{\sqrt{2}}(\gamma_0-i \gamma_x) \delta_0$ to write it in the following form
\begin{equation} \label{eq:SquaredHk}
\begin{split}
\tilde{\mathcal{H}}^2 =  & [3 v_1^2+9v_2^2+ \frac{v_p^2}{4} (k_x^2+k_y^2+k_z^2)] \gamma_0 \delta_0+ (2 v_1^2+6v_1 v_2) [\gamma_x \delta_0+ \gamma_x \delta_x + \gamma_0 \delta_x ] + \\
&  v_1 v_p [ k_z \gamma_z \delta_y - k_x \gamma_y \delta_0 + k_y \gamma_x \delta_y]+
3 v_2 v_p[k_y \gamma_0 \delta_y - k_x \gamma_y \sigma_x - k_z \gamma_y \sigma_z].
\end{split}
\end{equation}
In the eigenbasis, the matrix $\tilde{\mathcal{H}}^2$ has eigenvalues
\begin{equation} \label{eq:solvable_sol1}
 3 v_1^2+9v_2^2+ \frac{v_p^2}{4} (k_x^2+k_y^2+k_z^2) - 2 v_1 (v_1+3 v_2) \pm  v_p (v_1-3v_2) \sqrt{k_x^2+k_y^2+k_z^2} ,
\end{equation}
and
\begin{equation} \label{eq:d}
3 v_1^2+9v_2^2+ \frac{v_p^2}{4} (k_x^2+k_y^2+k_z^2) + 2 v_1 (v_1+3 v_2) \pm  (v_1+3v_2) \sqrt{16 v_1^2 + v_p^2 (k_x^2+k_y^2+k_z^2)}.
\end{equation}
Following Eq.~\eqref{eq:spectloc_squared}, the diagonal elements of $\tilde{\mathcal{H}}^2$, written in the eigenbasis,  
become part of the diagonal elements of $\mathcal{L}_k^2$. 
We focus first on Eq.~\eqref{eq:solvable_sol1} of
of matrix $\tilde{\mathcal{H}}^2$ which leads to the diagonal entries
\begin{equation} \label{eq:solvable_sol2}
- \kappa^2 \nabla_k^2 + \frac{v_p^2}{4} (k_x^2+k_y^2+k_z^2) \pm  v_p (v_1-3v_2) \sqrt{k_x^2+k_y^2+k_z^2} + 3 v_1^2+9v_2^2 - 2 v_1 (v_1+3 v_2)
\end{equation}
of $\mathcal{L}_k^2$.
Upon identifying $\kappa = \hbar, m= \frac{1}{2}$ and $\omega = v_p$, we recognize Eq.~\eqref{eq:solvable_sol2} as the Hamiltonian of a 3D quantum harmonic oscillator under a constant force and with a constant energy offset.
Using the spherical coordinates ($k_x = k \cos{\theta} \sin{\phi}, k_y = k \cos{\theta} \cos{\phi}$ and $k_z = k \sin{\theta}$), we obtain 
\begin{equation}\label{eq:ho_underforce}
H_{\rm h.o.} = -  \kappa^2 \nabla_k^2 + \frac{v_p^2}{4} k^2 \pm  v_p (v_1-3v_2) k  + v_1^2+9v_2^2 - 6 v_1 v_2,
\end{equation}
and by completing the square in $k$, Eq.~\eqref{eq:ho_underforce} becomes 
\begin{equation}
H_{\rm h.o.} = -  \kappa^2 \nabla_k^2  + \frac{v_p^2}{4} \left(k \pm \frac{2(v_1-3v_2)}{v_p}\right)^2.
\end{equation}
Upon replacing $\kappa^2 \rightarrow \frac{\hbar^2}{2m}, \frac{v_p^2}{4} \rightarrow \frac{m \omega^2}{2}, k' =k \pm \frac{2(v_1-3v_2)}{v_p}$ and $\nabla_k^2 \rightarrow \nabla_{k'}^2$, we recognize an equation of a 3D quantum harmonic oscillator in spherical coordinates.
Hence, for each diagonal element Eq.~\eqref{eq:solvable_sol1} of matrix $\tilde{\mathcal{H}}^2$, we obtain that matrix $\mathcal{L}_k^2$ has eigenvalues $ \epsilon^{\rm d, 1}_n = \tau_0 \sigma_0  \kappa v_p (n+\frac{3}{2})$.

We are left to calculate the diagonal elements of $\mathcal{L}^2_k$ corresponding to Eq.~\eqref{eq:d}, which read 
\begin{equation} \label{eq:ho_pert}
-\kappa^2 \nabla_k^2+ \frac{v_p^2}{4} (k_x^2+k_y^2+k_z^2)+ 3 v_1^2+9v_2^2 + 2 v_1 (v_1+3 v_2) \pm  (v_1+3v_2) \sqrt{16 v_1^2 + v_p^2 (k_x^2+k_y^2+k_z^2)}.
\end{equation}
In this case we are not able to find an general analytical expression, and so we 
resort to first-order perturbation theory in $v_1, v_2$ to determine the corresponding ground state energies.
For this, we separate two unperturbed $H_{\rm h.o.} = -\kappa^2 \nabla_k^2+ \frac{v_p^2}{4} (k_x^2+k_y^2+k_z^2)$ 
and the perturbation term $H_{\rm p}^{\pm}  = 3 v_1^2+9v_2^2 + 2 v_1 (v_1+3 v_2) \pm  (v_1+3v_2) \sqrt{16 v_1^2 + v_p^2 (k_x^2+k_y^2+k_z^2)}$, which contains all terms proportional to $v_1, v_2$.
In spherical coordinates, $H_{\rm h.o.} = -\kappa^2 \nabla_k^2+ \frac{v_p^2}{4} k^2$ and $H_{\rm p}^{\pm}  = 3 v_1^2+9v_2^2 + 2 v_1 (v_1+3 v_2) \pm  (v_1+3v_2) \sqrt{16 v_1^2 + v_p^2 k^2}$.
The ground state energy of $H_{\rm h.o.}$ is $\frac{3 \kappa v_p}{2}$ and the corresponding wave-function is 
$\Psi_{GS} (\mathbf{k}) = (\beta^2/\pi)^{3/4} \exp[-\beta^2 k^2/2]$ where $\beta = \sqrt{v_p/2\kappa}$.
In the first approximation, $H_{\rm p}^{\pm}  $ contributes with the energy $E_{\rm p}^{\pm}  = 4\pi \int_{0}^{\infty} \Psi^{*}_{GS} (\mathbf{k}) H_{\rm h.o.}  \Psi_{GS} (\mathbf{k})  k^2 dk $.
Here, we have already used the fact that $H_{\rm h.o.} $ is a spherically symmetric function such that the angular dependence is already integrated out. 
We solve this integral numerically.  
To simplify the analysis, we assume that parameters $v_1$ and $v_2$ are not independent. 
In particular $v_2 = v_1/\chi$ where $\chi = 3.4375$ is chosen such that it reflects the ratios of amplitudes $v_1$ and $v_2$ for crystalline RhSi. 
In Fig.~\ref{fig:qho_correction}, we plot the perturbed eigenvalues $ \epsilon^{\rm d, 2\pm}_0 = \frac{3 \kappa v_p}{2}+ E_{\rm p}^{\pm} $ as a function of parameter $v_1$.
We see that $H_{\rm p}^+$ monotonously increases the ground state energy while $H_{\rm p}^-$ mildly reduces it for $v_1 \lessapprox 1.9$.

Lastly, we evaluate the eigenvalues of the off-diagonal block of operator $\mathcal{L}_k^2$.
From Eqs.~\eqref{eq:spectloc_squared} and \eqref{eq:Hk}, we obtain 
\begin{equation}\label{eq:offdiag}
\kappa  \sum_{j=1}^d \sigma_j \partial_{k_j} \mathcal{H} = \frac{v_p \kappa}{2}  (\sigma_x  \gamma_z \delta_x+ \sigma_y  \gamma_0 \delta_y + \sigma_z \gamma_z \delta_z). 
\end{equation}
The eigenvalues of this off-diagonal part are not changed upon subsequent unitary rotations used to obtain matrix $\tilde{\mathcal{H}}$ and bring it into the eigenbasis. 
They consist of doubly degenerate $\epsilon^{\rm off-d, 1 \pm } = \pm \frac{3 v_p \kappa}{2}$ and six-fold degenerate $\epsilon^{\rm off-d, 2 \pm}  = \pm \frac{ v_p \kappa}{2}$. 
\end{widetext}

\begin{figure}[tb]
\includegraphics[width=8.6cm]{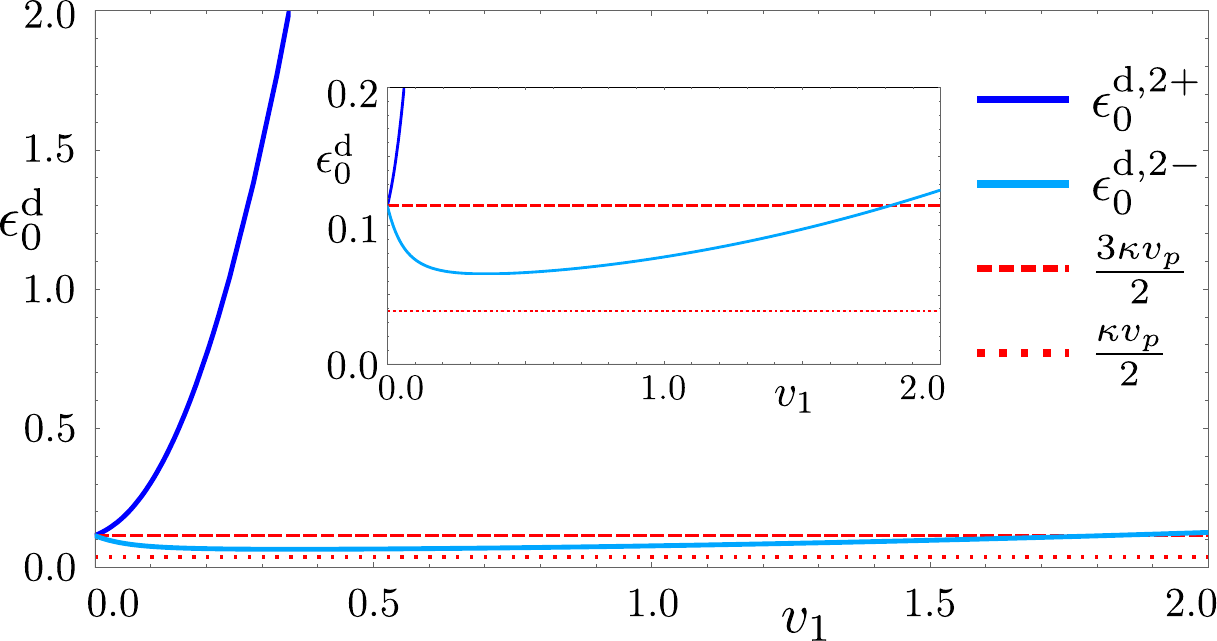}
\caption{
Ground state energy for the Hamiltonians Eq.~\eqref{eq:ho_pert} 
as function of $v_1$.
This energy is determined using first-order perturbation theory.
Dashed and dotted lines represent positive eigenvalues of the 
matrix Eq.~\eqref{eq:offdiag}.
}
\label{fig:qho_correction}
\end{figure}

Finally, we can determine the ground state energy of operator $\mathcal{L}_k^2$.
As before, it is a sum of diagonal and off-diagonal contributions.
Since the maximal amplitude of off-diagonal contributions is $\frac{3 v_p \kappa}{2}$, we focus on $n=0$ limit for the diagonal contributions.
These are eight-fold degenerate $\epsilon_0^{\rm d, 1} = \frac{3 \kappa v_p}{2}$ and four-fold degenerate $\epsilon_0^{\rm d, 2\pm} = \ \frac{3 \kappa v_p}{2}+ E_{\rm p}^{\pm} $ plotted in Fig.~\ref{fig:qho_correction}. 
Clearly, $ \epsilon_0^{\rm d, 1} +\epsilon^{\rm off-d, 1- } = 0 $ and is doubly degenerate because $\epsilon^{\rm off-d, 1 \pm } $ is doubly degenerate.
Moreover, $\epsilon_0^{\rm d, 1}  + \epsilon_0^{\rm off-d, 2\pm}$ is always nonzero.
Regarding $\epsilon^{\rm d, 2 \pm}  $, $\epsilon^{\rm d, 2 \pm}  +\epsilon^{\rm off-d, 1- }  =E_{\rm p}^{\pm}  $ that is nonzero $0< v_1 < 1.5$ as shown in Fig.~\ref{fig:qho_correction}.
This range of values covers the parameter regimes of RhSi and CoSi. 
From Fig.~\ref{fig:qho_correction}, we see that $\epsilon^{\rm d, 2 \pm} $ never crosses any off-diagonal contribution $\epsilon^{\rm off-d, 1 \pm }/\epsilon^{\rm off-d, 2 \pm }$ that is represented with dashed/dotted horizontal red lines.
Thus, for $0< v_1 < 1.5$ these diagonal terms can never produce zero states of operator $\mathcal{L}_k^2$. 
We conclude that the operator $\mathcal{L}_k^2$ has two zero modes for a threefold fermion of monopole charge $C = 2$ in case of RhSi/CoSi.
Due to a double-Weyl fermion at the $R$ point, $\mathcal{L}_k^2$ will have four zero states in total for parameter regime (2). 
This semi-analytical result is confirmed upon diagonalizing the full 
real space operators $\mathcal{L}^2$ and $\mathcal{L}$ for parameters $v_1 = 0.55,v_2 = 0.16, v_p = -0.762$, see Fig.~\ref{fig:figsm7}e, f.

In numerics, we observe that increasing values of $\kappa$ reduces the range of $v_1, v_2$ values for which our analytical prediction holds.
This is consistent with the expectation stemming from a more rigorous analysis of the eigenvalues of operator $\mathcal{L}_k^2$ 
in Eq.~\eqref{eq:spectloc_squared}, where $\kappa$ is necessarily assumed to be a small quantity~\cite{Schulz-Baldes2021, Schulz-Baldes2022}. 
Indeed, for large values of $\kappa$, we cannot ignore the tunnel effect between two harmonic wells that increases the energies of states within these wells~\cite{Schulz-Baldes2021}.
Hence, the analysis in this section is usually referred to as semi-classical~\cite{Schulz-Baldes2021, Schulz-Baldes2022}.

In addition, for smaller values of $\kappa \sim 0.1$, we obtain that increasing $v_1, v_2$ also leads to the breakdown of the semi-classical analysis. 
Such a behavior is expected because nonzero $v_1,v_2$ produce multifold fermions at $\Gamma$ and $R$ points at different energies. 
The energy difference in fact increases with larger $v_1, v_2$, implying that the spectral localizer calculated at $E=0$ is not anymore capturing only the low-energy physics of the problem.
The spectrum of operator $\mathcal{L}$ in this case can be derived by mapping it to the Dirac equation~\cite{Franca2023}.

\section{The effect of parameter $\kappa$ on the spectral localizer's spectrum}

\begin{figure*}[tb!]
\includegraphics[width=0.8\textwidth]{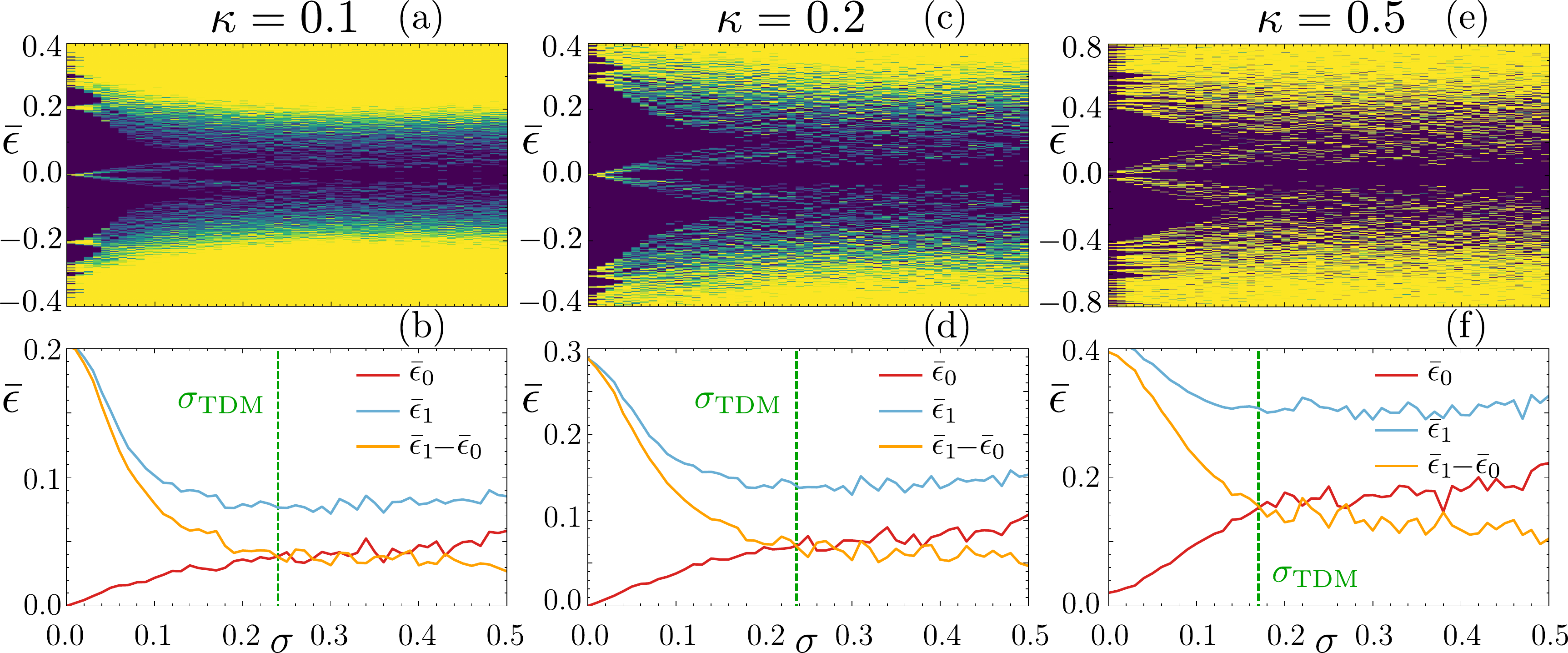}
\caption{
Panels (a), (c) and (e) show the disorder averaged density of states $\bar{\rho}_{\mathcal{L}_0}$ of operator $\mathcal{L}_0$ as a function of disorder for parameters $\kappa = 0.1,0.2$ and $0.5$, respectively.
Panels (b), (d) and (f) show the corresponding disorder averaged energy levels $\bar{\epsilon}_0, \bar{\epsilon}_1$ as well as their difference $\bar{\epsilon}_1-\bar{\epsilon}_0$.
Here, we consider $51$ equally spaced strengths of disorder in the window $\sigma \in [0,0.5]$ and $\rm N_{dis} = 10$ disorder realizations for every $\sigma$.
The spectral localizer captures the topology of a-RhSi. 
}
\label{fig:figsm11}
\end{figure*}

As we have seen in the previous section, the parameter $\kappa$ used to define the spectral localizer, see Eq.~\eqref{eq:spectloc}, determines the energetics of the harmonic oscillators obtained from the semi-classical analysis.
This analysis suggests that increasing $\kappa$ leads a larger gap between the ground state and the first excited state of this oscillator that would also translate into a larger bulk gap of the spectral localizer.
However, larger $\kappa$ increases the strength of tunelling between potential wells that host zero modes thus lifting their energy away from zero~\cite{Schulz-Baldes2021, Schulz-Baldes2022}.
While the tunneling effect causes a negligible split in energies, compared to the bulk gap in crystalline systems, it might have some unforeseen consequences in the presence of strong disorder.
Note that the analytical predictions of Refs.~\cite{Schulz-Baldes2021, Schulz-Baldes2022} concern the case of weak disorder that can be captured with first order perturbation theory. 
Refs.~\cite{Schulz-Baldes2021, Schulz-Baldes2022} also provide the analytical bound for parameter $\kappa$
\begin{equation}
\kappa < \frac{12 g^3}{ ||H || \; ||[D,H]||},
\end{equation}
where $g = 1/||H^{-1}||$ and $|| X||$ represents the norm of matrix $X$.
In practice, this bound requires unrealistically small values of parameter $\kappa$ for the considered systems and thus cannot be used.
Therefore, we choose the value of parameter $\kappa$ such that (I) the spectrum of the $\mathcal{L}$ operator contains four midgap modes and is otherwise gapped; and (2) the splitting of midgap modes due to the tunneling effect should be negligible.   
Hence, our criteria allow for a wide range of $\kappa$ values to be used, just like a large span of $\kappa$ values can be used to detect topological phases in Chern insulators~\cite{Loring2019}.
We showcase this by examining how the spectral localizer spectrum evolves with $\sigma$ for three values of parameter $\kappa = 0.1,0.2, 0.5$.
The results are shown in Fig.~\ref{fig:figsm11}.
As in the main text, we choose to evaluate the spectral localizer at position $\mathbf{r} = (0,0,0)$ and energy $E=0$, hence $\mathcal{L}(\mathbf{r},E) = \mathcal{L}_0$.

In Figs.~\ref{fig:figsm11}a,b, we repeat the calculation for $\kappa = 0.1$ with more disorder values in range $\sigma \in (0.2,0.5)$ compared to the main text.
We observe that the gap closing between $\bar{\epsilon}_0$ and $\bar{\epsilon}_1-\bar{\epsilon}_0$ occurs at $\sigma_{\rm TDM} = 0.25$ and is thus in agreement with the result of the main text.
For $\kappa = 0.2$, we see from Figs.~\ref{fig:figsm11}c,d that $\sigma_{\rm TDM} \approx 0.25$, i.e., the topological phase transition point remains invariant under the change of $\kappa$.
Note that the gap between midgap and first excited state in the crystalline limit is $\approx 1.5$ times larger for $\kappa = 0.2$ compared to $\kappa = 0.1$.
However, larger $\kappa$ contributes to the larger spread of midgap modes with disorder, resulting in the same topological phase transition point.
Finally, we consider the case of $\kappa = 0.5$.
As we can see from Figs.~\ref{fig:figsm11}e,f the size of the bulk gap is $\approx 2$ times larger for $\kappa = 0.5$ compared to $\kappa = 0.1$ in the crystalline limit.
However, such a large value of $\kappa$ causes the visible split in energies of midgap modes as well as a considerable spread in energies of midgap modes as disorder is increased.
We presume these effects are caused by the aforementioned tunelling between potential wells.
In total, it would seem that the operator $\mathcal{L}_0$ calculated with $\kappa = 0.5$ suggests a topological phase transition occurring at a smaller strength of disorder $\sigma \approx 0.17$.
This contradicts not only $\sigma_{\rm TDM}=0.25$ determined with $\kappa = 0.1,0.2$ but also the Fermi arcs considerations that are independent of operator $\mathcal{L}_0$, see Appendix~\ref{sec:Farcs}, and that indicate that the Fermi arcs survive up till $\sigma \lessapprox \sigma_{\rm TDM}$.
Due to the bulk-boundary correspondence, this provides a $\kappa$-independent proof that the bulk topology survives up till $\sigma_{\rm TDM}= 0.25$.
Hence, this would indicate that we should take values of $\kappa$ such that we do not observe a visible splitting of midgap modes in $\mathcal{L}_0$ spectrum.

\section{Results for CoSi}

In the main text, we have analyzed how amorphicity affects RhSi using different quantities such as the density of states, adjacent level spacing ratio and spectral localizer. 
In this section, we repeat those calculations for parameters chose to model a-CoSi.

\begin{figure}[tb]
\includegraphics[width=8.6cm]{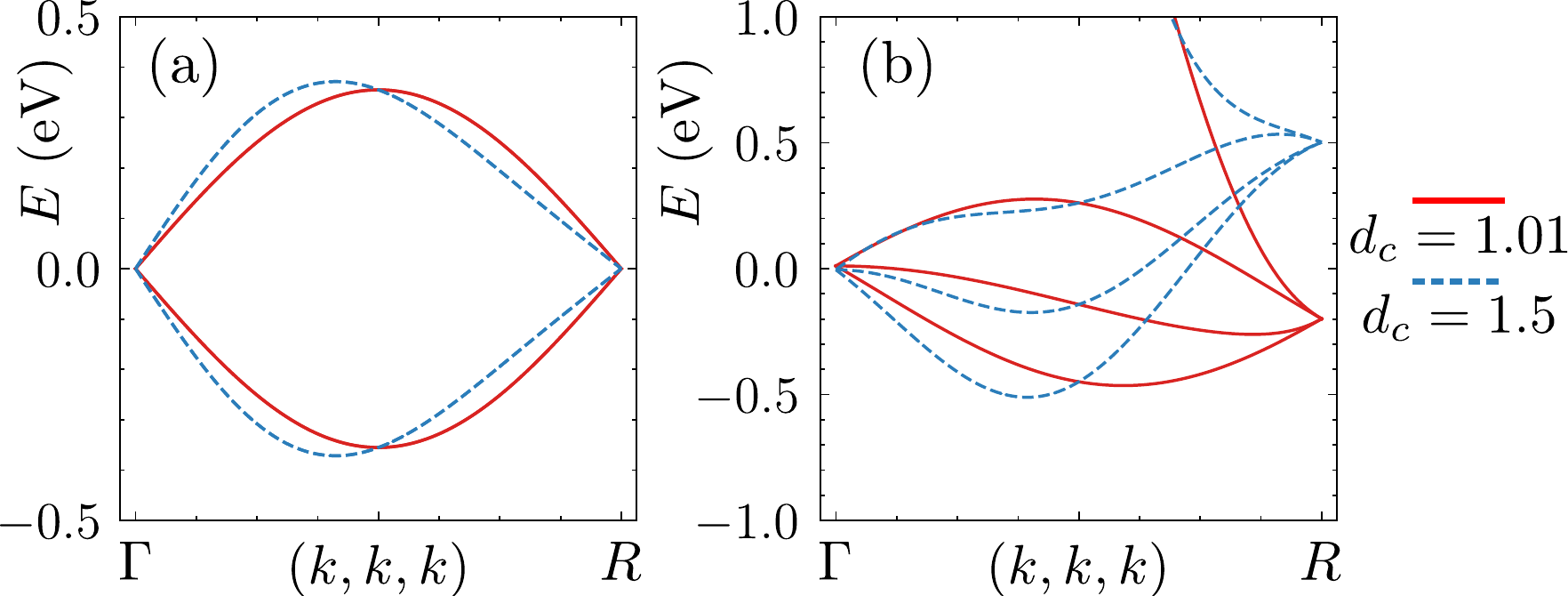}
\caption{
The band structure of crystalline CoSi in parameter regimes (1) and (2) is shown in panels (a) and (b), respectively. 
Red curves show the spectrum for cutoff distance $d _c = 1.01$, while blue curves show the spectrum once $d _c = 1.5$.
}
\label{fig:figsm3}
\end{figure}

Like in the main text, we consider two parameter regimes. 
In regime (1), parameters read $v_1=v_2= 0$ and $v_p = 0.41$.
The regime (2) has parameters $\mu = 0.551, v_1 = 1.29,v_2 = 0.25$ and $v_p = 0.41$, where $\mu$ is the chemical potential represented by $\mathcal{H}_0 = \mu \gamma_0 \delta_0$. 
Here, all parameters are given in units of eV.
Since parameter regime (2) reproduces well the band structure of crystalline CoSi at the Fermi level~\cite{Xu2020}, 
we shall often refer to it as the a-CoSi regime in the following.

The resulting band structures for regimes (1) and (2) are shown in Figs.~\ref{fig:figsm3}a and b, respectively.
Here, solid red lines represent the band structure corresponding to the Hamiltonian $\mathcal{H}_0+\mathcal{H}_1+\mathcal{H}_2$.
Dashed blue lines indicate the band structure for a cutoff distance $d_c = 1.5$ that corresponds to studying the Hamiltonian $\mathcal{H}_0+ \mathcal{H}_1+\mathcal{H}_2+\mathcal{H}_3+\mathcal{H}_4$. 
We see that the spectrum of CoSi in regime (1) along the $\Gamma-R$ path looks very similar to the corresponding spectrum of RhSi shown in Fig.~1c of the main text, in the sense that the threefold fermion is placed at higher energy compared to the double-Weyl fermion.
This is not anymore the case for regime (2), as we see that longer range hoppings lift the double-Weyl fermion higher in energy compared to the threefold fermion, unlike the situation in RhSi. 
Lastly, we see that bands along the $\Gamma-R$ path disperse less in CoSi than in RhSi.

\subsection{Spectral properties}

\begin{figure}[tb]
\includegraphics[width=8.6cm]{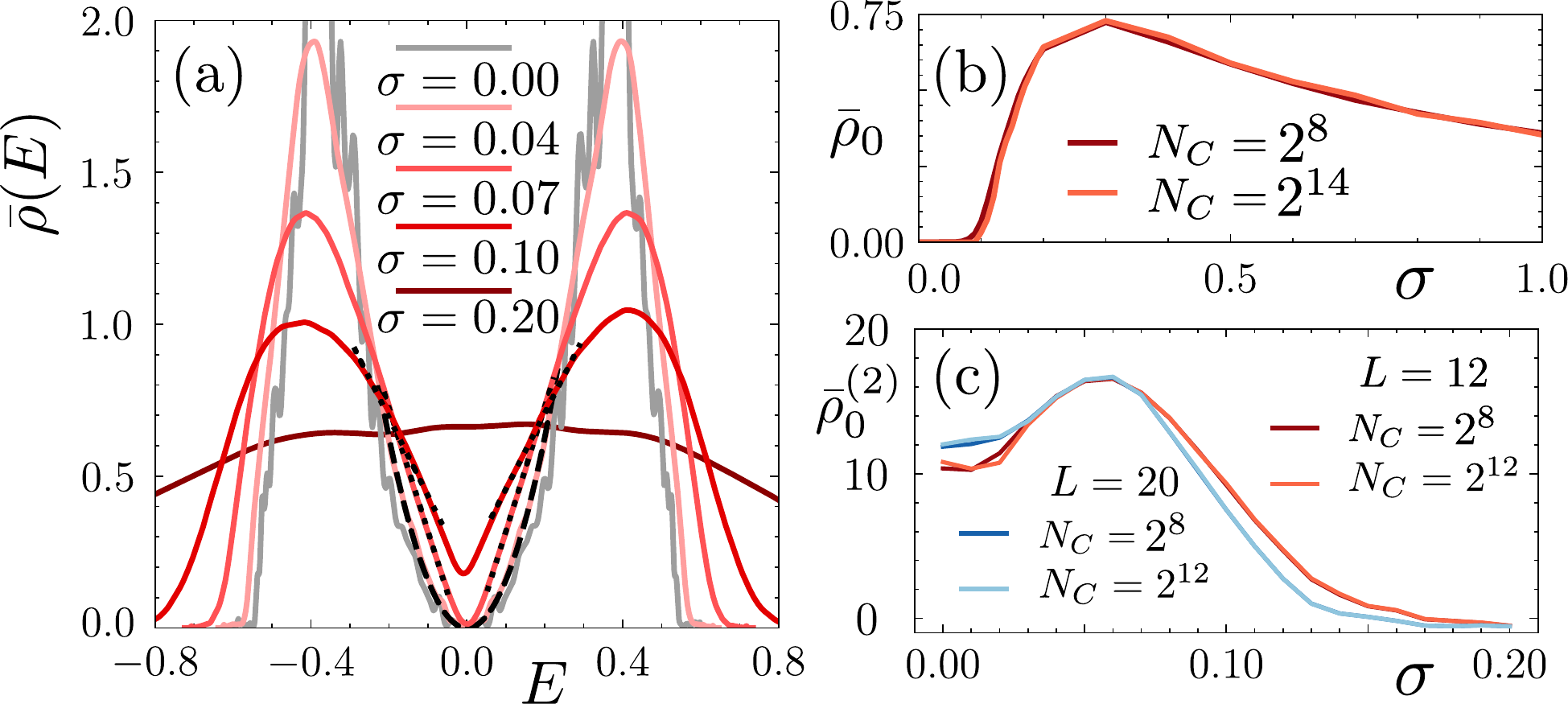}
\caption{
Panel (a) shows how $\bar{\rho}(E)$ vs $E$ changes with $\sigma$. Dashed and dotted lines represent fit functions $\alpha E^2$ and $\alpha + \beta |E|$, respectively. 
Panel (b) shows how zero-energy averaged DOS $\bar{\rho}_0$ changes with $\sigma$. 
Panel (c) shows $\bar{\rho}^{(2)}_0$ as a function of disorder strength. 
We define $\bar{\rho}^{(2)} (0)=\sum_{\lambda=1}^{N_{\rm dis}}(\rho^{\lambda})^{(2)} (0) $ where $(\rho^{\lambda})^{(2)} (0) $ is estimated from a fit $\rho^{\lambda}(E) = \rho^{\lambda}_0+ (\rho^{\lambda})^{(2)} (0) E^2$ 
in the energy range $(-0.2,0.2)$ for independent disorder realizations $\lambda$.
}
\label{fig:figsm4}
\end{figure}

Having fewer dispersive bands greatly affects spectral properties of a-CoSi, as it allows disorder to couple states at different energies with smaller energy penalty.  
Fig.~\ref{fig:figsm4}a shows the averaged density of states for a-CoSi in parameter regime (1). 
We immediately observe that $\bar{\rho} (E)$ for a crystalline system reaches its maximal values for approximately two times smaller energies compared to a-RhSi, see also Fig.~2a of the main text.
For small values of disorder ($\sigma \approx 0.04$) $\bar{\rho} (E) \propto E^2$, characteristic for linearly dispersing bands.
As disorder strength is increased to $\sigma \approx 0.07$, a behavior $\bar{\rho} (E) \propto |E|$ emerges that is characteristic of the proximity of the quantum critical point. 
This behavior persists for $\sigma = 0.1$, along with nonzero DOS at zero energy. 
Finally, stronger disorder strengths yield broad and featureless DOS characteristic for diffusive metals.
In Fig.~\ref{fig:figsm4}b, we plot $\bar{\rho}_0$ as a function of disorder strength,
and see it becomes nonzero for $\sigma \lessapprox 0.1$.

To study the semimetal-diffusive metal phase transition, we plot in Fig.~\ref{fig:figsm4}c the second derivative $\bar{\rho}_0^{(2)}$ as a function of disorder for different system sizes and orders $N_C$ of the KPM expansion. 
This quantity peaks at $\sigma_{c} \approx 0.06$ signaling the proximity of the quantum critical point.
It however does not diverge with varying parameters, implying that the phase transition between semimetal and diffusive metal regimes is avoided, just like for a-RhSi in parameter regime (1).
%

 \subsection{Anderson localization}

\begin{figure}[tb]
\includegraphics[width=8.6cm]{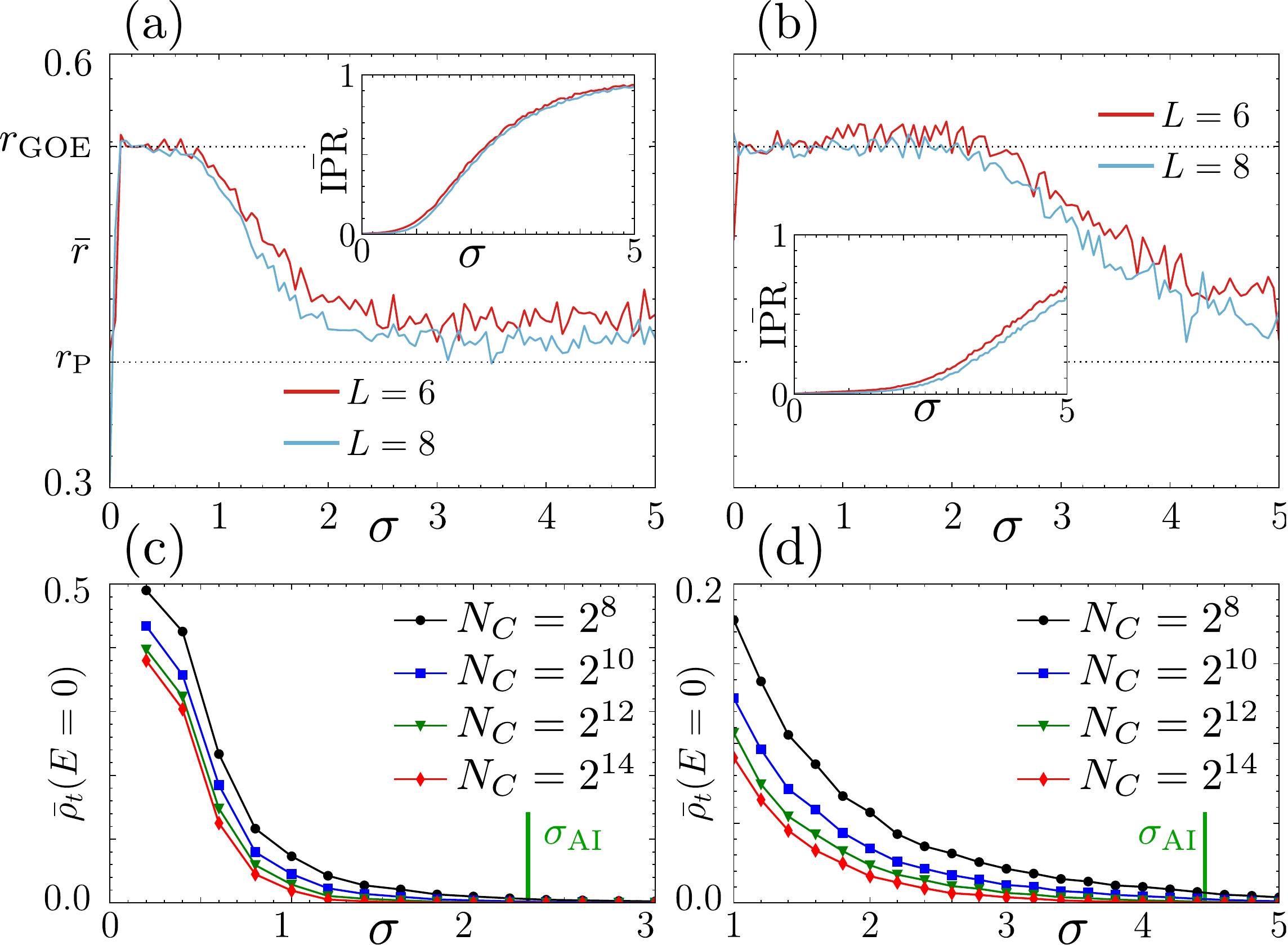}
\caption{
Panels (a) and (b) show adjacent level spacing ratios at $E_F = 0$ as a function of disorder strength $\sigma$ for parameter regimes (1) and (2), respectively. 
Here, $r_{\rm GOE} = 0.536$ and $r_{\rm P} = 0.387$ are represented by dotted gray lines. 
The insets show how the corresponding inverse participation ratios at the Fermi energy evolve with $\sigma$.
Panels (c) and (d) show the disorder averaged typical DOS for different orders $N_C$ of the KPM expansion as a function of $\sigma$ for a-CoSi in regime (1) and (2), respectively.
We calculate $\bar{\rho}_t$ using $N_{\rm dis} = 51$ disorder realizations for a system of linear size $L=20$, and we take $V_t = 32$, like in the main text.
Here, $\sigma_{\rm AI}$ (green line) is the critical disorder strength at
which the system transitions from a DM to an AI phase.
}
\label{fig:figsm5}
\end{figure}

To study localization properties of a-CoSi, we use the disorder averaged adjacent level spacing ratio $\bar{r}$ and inverse participation ratio $\rm \bar{IPR}$ at the Fermi level.
The results for parameter regime (1) and (2) are shown in Figs.~\ref{fig:figsm5}a and b, respectively.
From Fig.~\ref{fig:figsm5}a, we see that the system in regime (1) exhibits a diffusive metal behavior for $\sigma \lessapprox 1$, after which $\bar{r}$ gradually decays from $r_{\rm GOE}$ to $r_{\rm AI}$. 
Therefore, the range of disorder strengths for which the system exhibits a typical diffusive metal behavior is smaller for a-CoSi in regime (1) compared to a-RhSi in the same regime. 
We can understand this as a consequence of smaller hopping amplitudes that imply a larger DOS near zero energy and thus a faster localization of states.
In the a-CoSi regime (2), we see from Fig.~\ref{fig:figsm5}b that the diffusive metal phase persists to a much larger disorder strength $ \sigma \approx 2.6$, compared to regime (1).
In fact, the a-CoSi system in regime (2) seems more robust to Anderson localization than a-RhSi in the same regime, see Fig.~3 of the main text.
This is because the amorphous system supports additional hoppings that make the energy splitting of threefold and double-Weyl fermions larger in a-CoSi than for a-RhSi.

To determine the disorder strength at which the Anderson localization occurs, we calculate the typical DOS for different KPM expansion orders as explained in the main text. 
The results are shown in Figs.~\ref{fig:figsm5}c and d for parameter regimes (1) and (2) of a-CoSi, respectively.
Using Richardson extrapolation, we obtain $\sigma_{\rm AI} = 2.40 $ for regime (1), and $\sigma_{\rm AI} = 4.46$ for regime (2).

\subsection{Topological phase diagram}

\begin{figure}[tb]
\includegraphics[width=8.6cm]{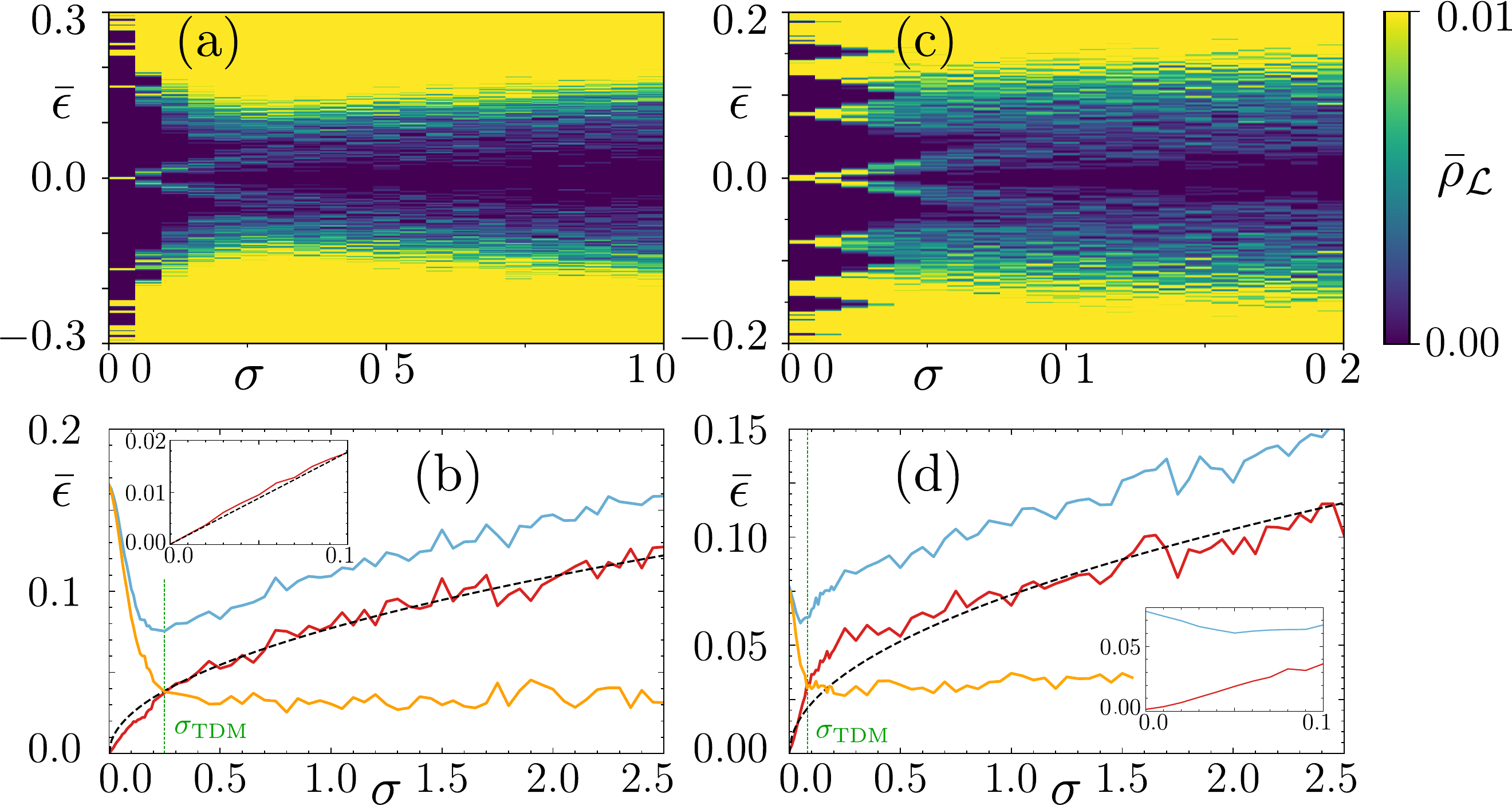}
\caption{
Panels (a) and (b) show disorder averaged DOS of the operator $\mathcal{L}_0$ as a 
function of disorder strength $\sigma$ for parameter regimes (1) and (2), respectively. 
In panels (c) and (d) are plotted disorder averaged energies of midgap and first-excited states 
$\bar{\epsilon}_0, \bar{\epsilon}_1$, and their difference $\bar{\epsilon}_1-\bar{\epsilon}_0$, as a function of $\sigma$ for these two parameter regimes. 
We represent $\bar{\epsilon}_0$, $\bar{\epsilon}_1$ and $\bar{\epsilon}_1-\bar{\epsilon}_0$ with red, blue and orange color, respectively.
A dashed green line represent the topological phase transition point $\sigma_{\rm TDM}$.
Dashed black lines represents a fit $\bar{\epsilon}_0 = a \sqrt{\sigma}$, where $a =0.077$ in  regime (1) and $a= 0.073$ for regime (2). 
The inset of panels (b) and (d) show how well $\bar{\epsilon}_0$ matches with the predicted form $\kappa^{0.75} \sigma$~\cite{Schulz-Baldes2022} (gray line) in case of small disorder strengths. 
Here, we use $\kappa = 0.1, \rm{N_{dis} }= 25$ and $L = 12$. 
}
\label{fig:figsm6}
\end{figure}

Lastly, we review the topological phase diagram of a-CoSi.
The results are shown in Fig.~\ref{fig:figsm6}.
To produce this figure, we used $\rm N_{dis} = 25$ disorder realizations and cubic systems of linear size $L= 12$ unit cells.  

For parameter regime (1), we observe that the topological phase diagram shown in Figs.~\ref{fig:figsm6}a,b looks very similar
to the phase diagrams calculated for a-RhSi in both regimes.
In addition to having a vanishing gap between midgap and excited states in the DOS for $\sigma \gtrapprox 0.25$, we observe that $\bar{\epsilon}_1 = 2 \bar{\epsilon}_0$ for $\sigma \approx 0.25$ thus indicating a topological phase transition. 
In addition, we observe from the inset of Fig.~\ref{fig:figsm6}b that for low disorder strengths $\sigma \leq 0.1$, the averaged energy $\bar{\epsilon}_0$ of the $\mathcal{L}$ midgap mode follows the analytically predicted behavior $\bar{\epsilon}_0 = \sigma^{3/4} \kappa$~\cite{Schulz-Baldes2022}.
We also find that the dependence of $\bar{\epsilon}_0$ on $\sigma$ can be well approximated in the entire range $\sigma \in [0,2.5]$ by an $a \sqrt{\sigma}$ law, where $a = 0.077$.

In a-CoSi regime, the phase diagram in Fig.~\ref{fig:figsm6}c reveals that the topological properties of a-CoSi 
are less robust to effects of disorder compared to a-RhSi. 
The gap between $\bar{\epsilon}_0$ and $\bar{\epsilon}_1-\bar{\epsilon}_0$ eigenvalues of the $\mathcal{L}_0$ operator closes at $\sigma_{\rm TDM} \approx 0.08$, while in a-RhSi it closes for $\sigma_{\rm TDM} \approx 0.25$.
Moreover, for a crystalline system this gap is approximately two times smaller than the corresponding gap in regime (1) of the same material.
Disorder averaged eigenvalues $\bar{\epsilon}_0, \bar{\epsilon}_1$ shown in Fig.~\ref{fig:figsm6}d reveal that $\bar{\epsilon}_0$ is well approximated by $a \sqrt{\sigma}$ ($a = 0.073$) only for $\sigma >0.5$. 
The inset of Fig.~\ref{fig:figsm6}d shows $\bar{\epsilon}_0, \bar{\epsilon}_1$ for very small disorder strengths. 
We see that $ \bar{\epsilon}_1$ experiences an upturn at $\sigma \approx 0.075$, while  $\bar{\epsilon}_0$ cannot be well approximated with the analytical prediction $\bar{\epsilon}_0 =\sigma^{3/4} \kappa $, where $\kappa = 0.1$.

As mentioned in the main text, we relate this reduced topological robustness of a-CoSi compared to a-RhSi,
to different ratios of inter-orbital hoppings for the two cases.
As shown in Fig.~1 of the main text, there are two types of inter-orbital hoppings $v_1$ and $v_p$.
The latter one changes with the bond orientation, such that the hoppings $\frac{1}{4} (v_1\pm v_p)$ are arranged into the chiral structure resulting in the appearance of multifold fermions in parameter regime (2). 
In the presence of structural disorder that alters hoppings between orbitals, this chiral structure is gradually lost as $\sigma$ is increased. 
Thus, the relevant energy scale at which disorder effects alter topology is proportional to $\Delta v_{\rm nn} = \frac{v_p}{2} = \frac{1}{4} (v_1+ v_p)-\frac{1}{4} (v_1- v_p)$. 
 
\begin{figure}[tb!]
\includegraphics[width=8.6cm]{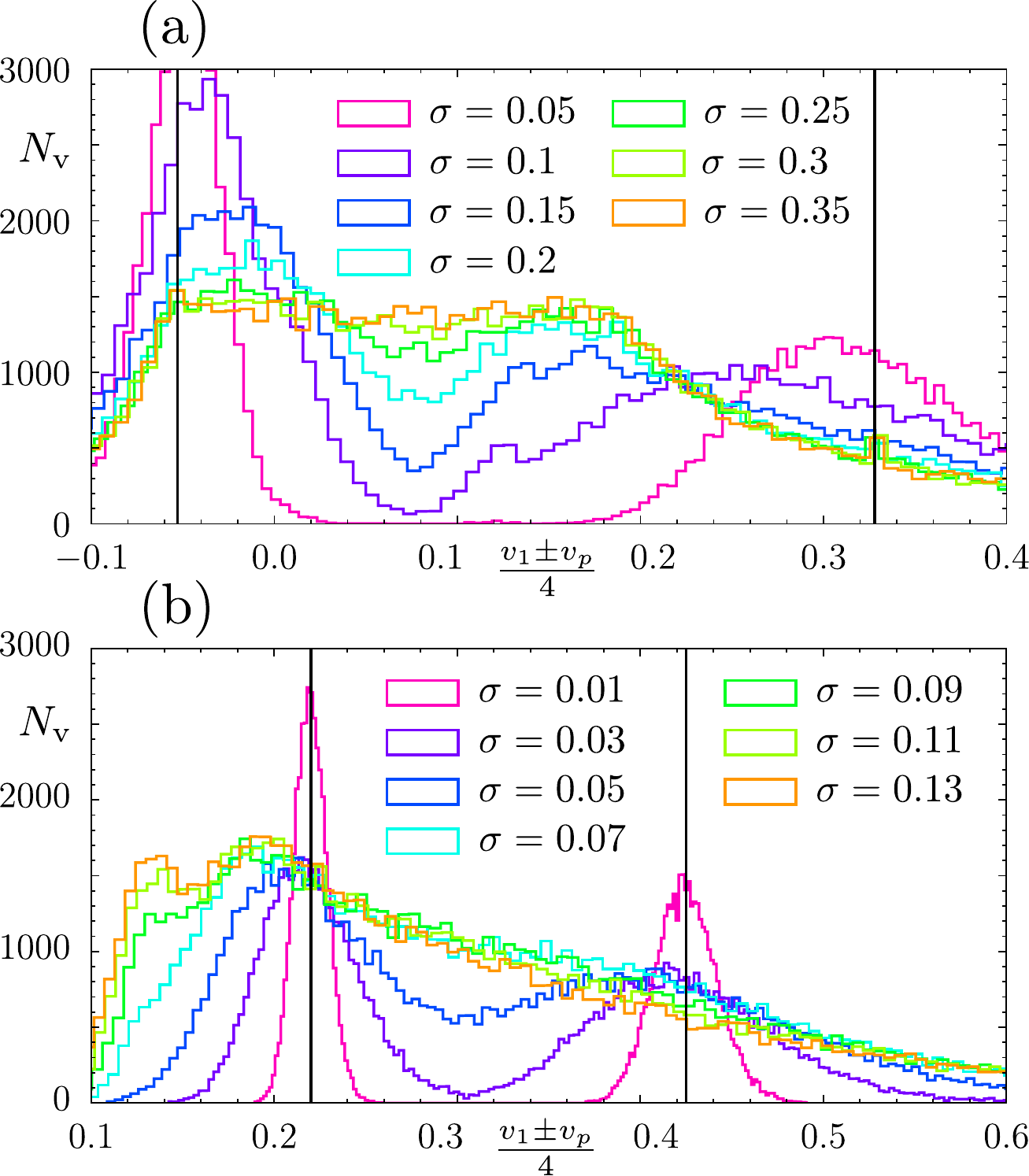}
\caption{
Panels (a) and (b) show the evolution of the histogram of nearest-neighbor hoppings in a-RhSi and a-CoSi with disorder. 
Vertical black lines correspond to the crystalline limit.
Here, we consider a system of linear size $L = 10$ and disorder realization $\lambda = 2$.
}
\label{fig:figsm8}
\end{figure}

In crystalline RhSi, $v_1 = 0.55$ and $v_p = -0.762$ such that the nearest-neighbor hoppings have amplitudes $\frac{1}{4} (v_1+v_p) = -0.053$ and $\frac{1}{4} (v_1-v_p) = 0.328$, and they differ by $|\Delta v_{\rm nn}|  = \frac{|v_2|}{2} = 0.381$.
For crystalline CoSi, parameters read $v_1 = 1.29$ and $v_p = 0.41$ implying that $\frac{1}{4} (v_1+v_p) = 0.425$, 
$\frac{1}{4} (v_1- v_p) = 0.22$ and their difference reads $|\Delta v_{\rm nn}|  = 0.205$.
To see how disorder changes these hoppings, we plot the histogram of all nearest-neighbor hoppings for a-RhSi and a-CoSi in Figs.~\ref{fig:figsm8}a and b, respectively.
We observe that the spread of these hoppings increases with $\sigma$, and eventually it is not possible to distinguish between two amplitudes of hoppings.
For a-RhSi this occurs for $\sigma \approx 0.3$, while for a-CoSi the relevant disorder strength is $\sigma \approx 0.09$.
Both of these values agree well with the topological phase diagram obtained with the spectral localizer.

Finally, in Table~\eqref{table:2}, we summarize all critical disorder strengths determined in this work for a-RhSi and a-CoSi in regime (2).

\begin{center}
\begin{tabular*}{\columnwidth}{@{\extracolsep{\fill}}|l|c|r|}
 \hline
quantities & a-RhSi & a-CoSi  \\ [0.5ex] 
 \hline\hline
$|\Delta v_{\rm nn} |$  & 0.328  & 0.205 \\ 
 \hline
$\sigma_{\rm TDM}$  & 0.25 & 0.08  \\
 \hline
$\sigma_{\rm AI}$ &  $2.98$ & $4.46$  \\ [1ex] 
 \hline
\end{tabular*}
\label{table:2}
\end{center}

\section{Evolution of Fermi arcs with disorder}\label{sec:Farcs}

In this section, we show how the Fermi arcs in a-RhSi and a-CoSi evolve with disorder. 
As explained in the main text, we capture the Fermi arcs using the disorder averaged momentum-resolved spectral function $\bar{\rm A} (\mathbf{k}, E) = \sum_{\lambda = 1}^{\rm N_{dis}} A^{\lambda} (\mathbf{k}, E) $ with $\rm N_{dis} = 10$. 
The function $\rm A^{\lambda} (\mathbf{k}, E)$, defined in Eq.~\eqref{eq:spect_func}, is calculated using the plane wave states $\ket{\mathbf{k}_o} = \exp[i (k_x x_n + k_y y_n)] \ket{n,o}$ where $x_n, y_n$ describe the position of site $n$ and orbital type $o$ in \textit{xy} plane located at the top layer of a cubic system. 
For a-RhSi, we look for the Fermi arcs at $E=-0.3$ while for a-CoSi we choose $E = -0.05$.  
Lastly, each of the functions $\rm A^{\lambda} (\mathbf{k}, E)$ is calculated using the KPM with $N_C = 256$.

\begin{figure*}[tb!]
\includegraphics[width=0.8\textwidth]{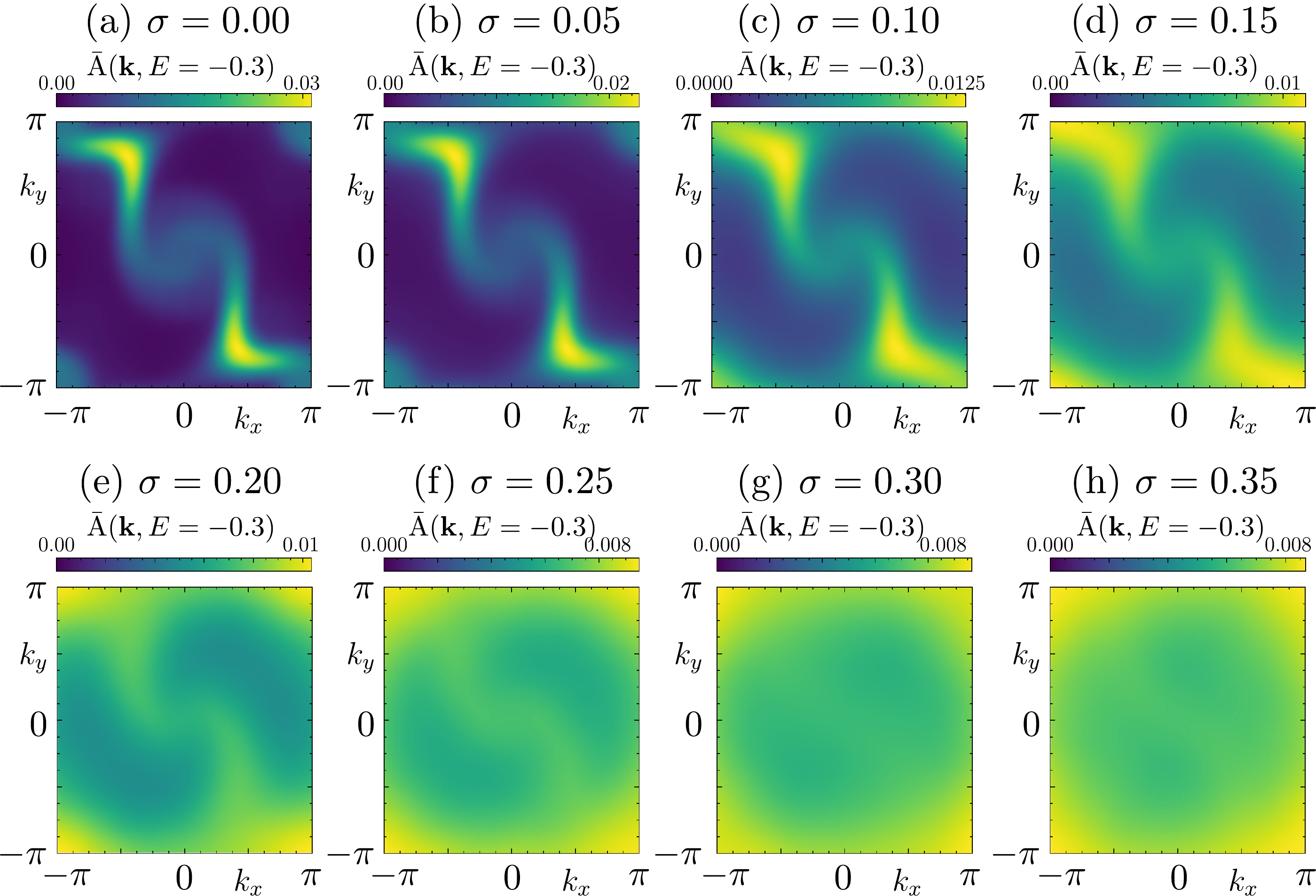}
\caption{
Panels (a)-(h) show the disorder averaged momentum-resolved spectral function $\bar{\rm A}(\mathbf{k}, E = -0.3)$ for different disorder strengths $\sigma$ for the a-RhSi.  
Here, we consider $N_{\rm dis} = 10$ disorder realizations of the system with $L = 12$ unit cells in each direction, and with parameters $v_1 = 0.55, v_2 = 0.16$ and $v_p = -0.762$. 
}
\label{fig:figsm9}
\end{figure*}

\begin{figure*}[tb!]
\includegraphics[width=0.8\textwidth]{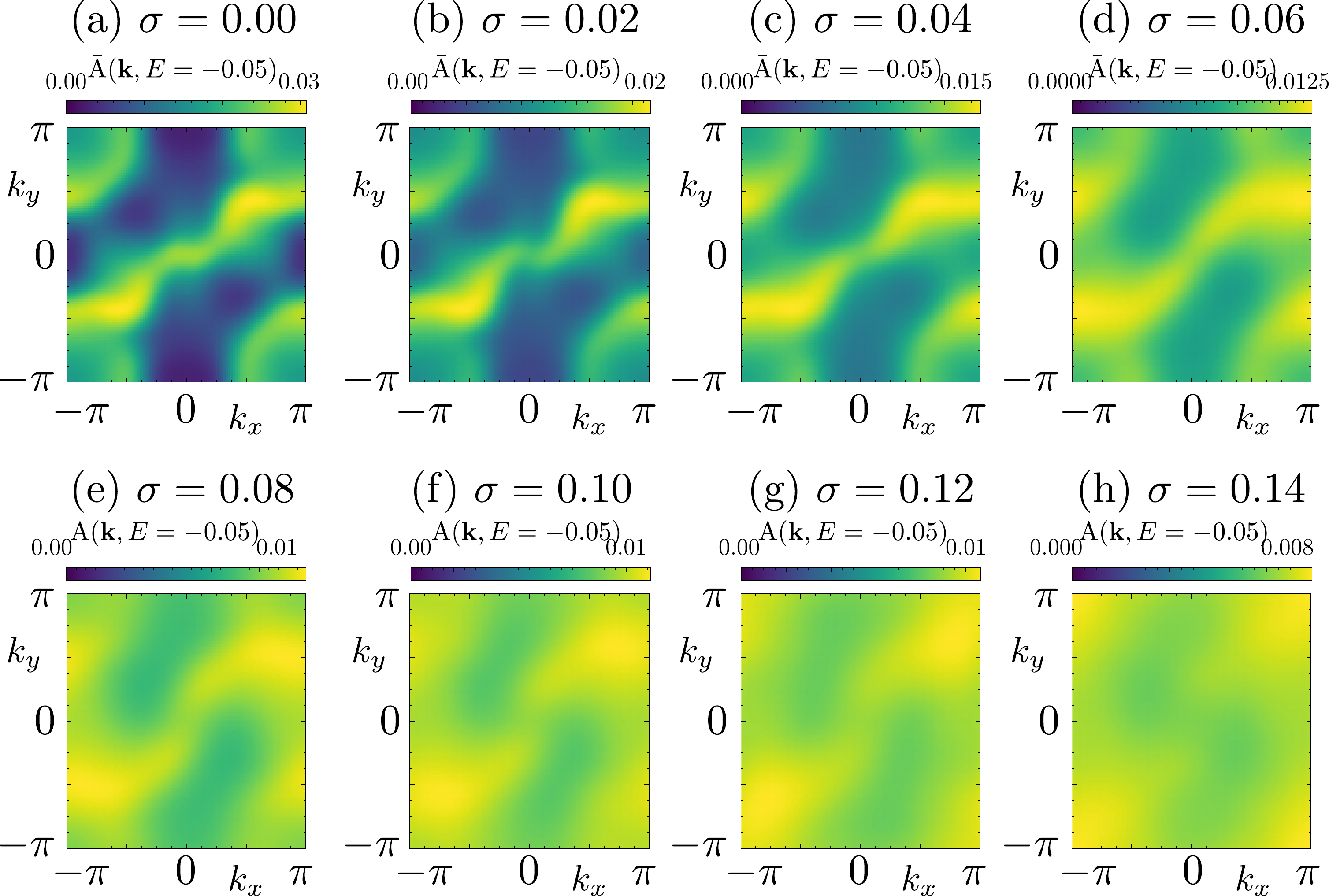}
\caption{
Panels (a)-(h) show the disorder averaged momentum-resolved spectral function $\bar{\rm A}(\mathbf{k}, E = -0.05)$ for different disorder strengths $\sigma$ for the a-CoSi.  
Here, we consider $N_{\rm dis} = 10$ disorder realizations of the system with $L = 12$ unit cells in each direction, and with parameters $\mu = 0.551, v_1 = 1.29, v_2 = 0.25$ and $v_p = 0.41$. 
}
\label{fig:figsm10}
\end{figure*}

The results are shown in Fig.~\ref{fig:figsm9} for a-RhSi and Fig.~\ref{fig:figsm10} for a-CoSi.
In both cases, we see that increasing disorder strength redistributes the spectral weight away from the Fermi arcs thus making the spectral function rotationally invariant in the momentum space. 
As explained in Ref.~\cite{Marsal2023}, this rotational invariance originates from the random orientation of bonds that are characteristic for amorphous systems.
Crucially, we observe that the disappearance of Fermi arcs occurs at approximately the same disorder strengths as $\sigma_{\rm TDM}$ estimated using the spectral localizer.

\end{document}